\documentclass[%
 aip,
 amsmath,amssymb,
reprint,%
]{revtex4-1}

\usepackage{graphicx}
\usepackage{dcolumn}
\usepackage{bm}

\usepackage[utf8]{inputenc}
\usepackage[T1]{fontenc}
\usepackage{mathptmx}

\newcommand{\norm}[1]{\left\lVert#1\right\rVert}

\usepackage[disable]{todonotes} 				

\newcommand*{\blauw}[1]{\textcolor{blue}{#1}}

\usepackage{amsmath,amssymb}
\usepackage{mathrsfs}
\usepackage{mathtools}
\usepackage{siunitx}
\usepackage{graphicx}
\usepackage{color}
\usepackage{dcolumn}
\usepackage{hyperref}
\usepackage[normalem]{ulem}
\usepackage{algorithm}
\usepackage{extarrows}
\usepackage[caption=false]{subfig}
\usepackage{soul}
\usepackage{algpseudocode}
\usepackage{comment}
\usepackage{mhchem}

\begin{document}

\preprint{AIP/123-QED}

\title[Sample title]{Sample Title:\\with Forced Linebreak}

\title{Reconstruction of Protein Structures from Single-Molecule Time Series}

\author{Maximilian Topel}
\affiliation{Pritzker School of Molecular Engineering, University of Chicago, Chicago, Illinois 60637}

\author{Andrew L. Ferguson}
\email{Author to whom correspondence should be addressed: \mbox{andrewferguson@uchicago.edu}}
\affiliation{Pritzker School of Molecular Engineering, University of Chicago, Chicago, Illinois 60637}

\date{\today}

\begin{abstract}
\noindent Single-molecule experimental techniques track the real-time dynamics of molecules by recording a small number of experimental observables. Following these observables provides a coarse-grained, low-dimensional representation of the conformational dynamics but does not furnish an atomistic representation of the instantaneous molecular structure. Takens' Delay Embedding Theorem asserts that, under quite general conditions, these low-dimensional time series can contain sufficient information to reconstruct the full molecular configuration of the system up to an \textit{a priori} unknown transformation. By combining Takens' Theorem with tools from statistical thermodynamics, manifold learning, artificial neural networks, and rigid graph theory, we establish an approach Single-molecule TAkens Reconstruction (STAR) to learn this transformation and reconstruct molecular configurations from time series in experimentally-measurable observables such as intramolecular distances accessible to single molecule F\"orster resonance energy transfer. We demonstrate the approach in applications to molecular dynamics simulations of a \ce{C_{24}H_{50}} polymer chain and the artificial mini-protein Chignolin. The trained models reconstruct molecular configurations from synthetic time series data in the head-to-tail molecular distances with atomistic root mean squared deviation accuracies better than 0.2 nm. This work demonstrates that it is possible to accurately reconstruct protein structures from time series in experimentally-measurable observables and establishes the theoretical and algorithmic foundations to do so in applications to real experimental data.
\end{abstract}

\maketitle

\section{Introduction} \label{sec:intro}

The molecular structure of a protein is defined by a vector in $\mathbb{R}^{3N}$ specifying the Cartesian coordinates of the $N$ constituent atoms. Molecular dynamics (MD) simulation is a computational algorithm that propagates the $3N$-dimensional position vectors through time under the action of a force field specifying the interatomic interaction potential \cite{frenkel2001understanding}. At each step of the MD simulation the full $3N$-dimensional configurational state of the system is available. Sophisticated experimental techniques such as X-ray crystallography and cryo-electron microscopy can solve protein structures to near atomic resolution in crystalline or vitrified samples \cite{PRODRG,Chang2011,Roy2008rr}. Proteins, however, are typically not functional under these conditions and these structures cannot capture transitions between metastable states. For example, the \textit{Mycobacterium tuberculosis} protein tyrosine phosphatase PtpB exhibits dynamical transitions between ``closed'' and ``open'' states in which a pair of $\alpha$-helices transiently cover the active site and dynamically protects the active site from oxidative inactivation \cite{flynn2010dynamic}. Single-molecule experimental techniques such as single molecule F\"orster resonance energy transfer (smFRET) can follow protein dynamics by tracking small numbers of experimentally observable intramolecular distances between fluorescent probes conjugated to the target molecule \cite{Chang2011,Roy2008rr,Zerze2014mo}. There are currently no experimental techniques available to follow the dynamical evolution in atomistic detail.

Takens' Delay Embedding Theorem is a result from dynamical systems theory that asserts a sufficiently long and frequently sampled time series of a single system observable can contain sufficient information to reconstruct the full-dimensional state of the system up to an \textit{a priori} unknown diffeomorphism (i.e., smooth and invertible bijective transformation) \cite{Takens,sauer1991embedology,packard1980,Broomhead1986,Cao1998,stark2003delay,complex1,holger,PMID:25733874}. This theoretical result opens the door to reconstructing the molecular coordinates of a protein from single-molecule experimental measurements such as smFRET. We have previously applied Takens' Theorem to synthetic single-molecule time series extracted from molecular dynamics simulations of polymers and proteins to estimate single molecule free energy surfaces (smFES) \cite{Ferg16,Ferg18}. Since the simulations also furnish the full molecular configurations we also estimated the ``true'' smFES from the atomistic molecular simulation trajectory to numerically verify the existence of the \textit{a priori} unknown diffemorphism and place empirical bounds on the degree of perturbation to the true smFES induced by this transformation. In this work, we build upon these foundations using tools from rigid graph theory and artificial neural networks to learn this transformation from the data and also approximate the inverse transformation from the low-dimensional smFES back up to the molecular configuration space. This calibrates a functional approximation mapping a time series in a single system observable to the atomistic molecular configuration, thereby enabling reconstruction of atomistic molecular structures from experimentally-measurable observables. We term this approach Single-molecule TAkens Reconstruction (STAR).

The structure of this paper is as follows. In the next section we describe the methodological details of STAR. We detail the mathematical formulation and numerical solution of each step of the learning problem combining principles and tools from statistical thermodynamics, manifold learning, and rigid graph theory. In Section \ref{sec:results} we present applications of STAR to MD simulations of a \ce{C_{24}H_{50}} polymer chain and the 10-residue artificial mini-protein Chignolin. We train STAR to reconstruct molecular configurations from univariate time series in the head-to-tail distance as a synthetic and idealized smFRET time trace. We demonstrate reconstruction of molecular configurations from novel time series data not present in the training ensemble to a root mean squared deviation (RMSD) accuracy better than 0.2 nm. In Section \ref{sec:concl} we present our conclusions and opportunities for future work.

\section{Methods} \label{sec:meth}

\subsection{Principles of STAR} \label{subsec:STAR}

\begin{figure*}[t]
    \centering
    \includegraphics[height=11cm, width=16cm]{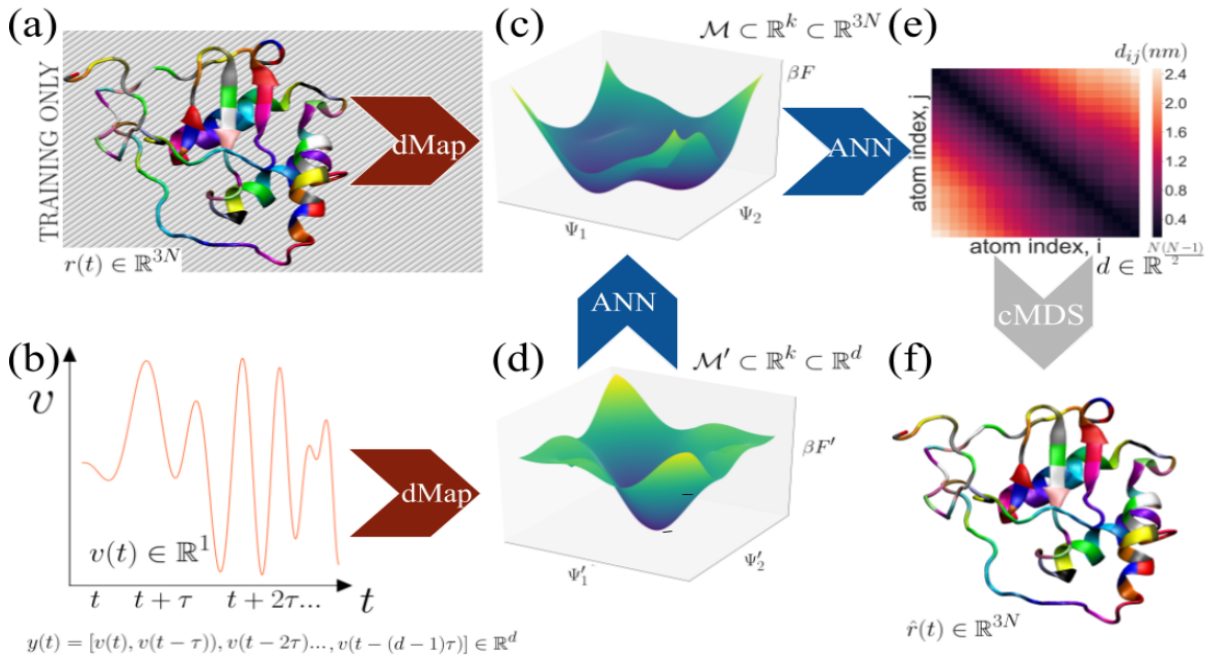}
    \linespread{1.0}\caption{Cartoon schematic of Single-molecule TAkens Reconstruction (STAR). The trained STAR model predicts molecular configurations of a protein $\hat{\mathbf{r}}(t) \in \mathbb{R}^{3N}$ from univariate time series in an experimentally-measurable observable of the system $v(t) \in \mathbb{R}^1$ through the four-step pathway $b \rightarrow d \rightarrow c \rightarrow e \rightarrow f$. Each panel corresponds to a different representation of the molecular system and arrows between them to mathematical operations. Red arrows represent unsupervised nonlinear manifold learning problems (i.e., nonlinear dimensionality reduction), blue arrows represent supervised learning problems (i.e., function approximation), and grey arrows represent deterministic operations. (a,b) Molecular dynamics simulations sufficiently long to sample the thermally-relevant configurational space provide training trajectories of molecular configurations $\mathbf{r}(t) \in \mathbb{R}^{3N}$ (panel a) and a scalar time series in an experimentally-measurable system observable $v(t) \in \mathbb{R}^1$ (panel b). In this work, we set $v(t)$ to be the head-to-tail ($h2t$) intramolecular distance that is, in principle, measurable by smFRET. (c) Interactions between atoms constrain the molecular trajectory $\mathbf{r}(t) \in \mathbb{R}^{3N}$ to a $k$-dimensional manifold $\mathcal{M} \subset \mathbb{R}^{k} \subset \mathbb{R}^{3N}$ with effective dimensionality $k \ll 3N$. The collective variables $\{\psi_1, \psi_2, \ldots \psi_k\}$ parameterizing the manifold are extracted from the simulation trajectory using diffusion maps. The manifold $\mathcal{M}$ supports the smFES $\beta F(\psi_1, \psi_2, \ldots \psi_k)$. (d) An image of $\mathcal{M}$ and the smFES is obtained from the time series data $v(t) \in \mathbb{R}^1$ by forming a $d$-dimensional delay embedding $\mathbf{y}(t) = [v(t), v(t-\tau), v(t-2\tau), \ldots , v(t-(d-1)\tau)] \in \mathbb{R}^d$ and then applying diffusion maps to learn a parameterization $\{\psi_1^\prime, \psi_2^\prime, \ldots \psi_k^\prime\}$ of a manifold $\mathcal{M}^\prime \subset \mathbb{R}^{k} \subset \mathbb{R}^{d}$. Under technical conditions on $v$, $d$, and $\tau$ discussed in the main text, Takens' Delay Embedding Theorem asserts that the dynamical evolution of $\mathbf{y}(t)$ is $C^1$-equivalent to that of $\mathbf{r}(t)$ and that the manifold $\mathcal{M}^\prime$ is related to $\mathcal{M}$ via diffeomorphic (i.e., smooth, invertible, and bijective) transformation $\Theta : \mathcal{M}^\prime \rightarrow \mathcal{M}$. We learn the transformation $\Theta$ from the training data using a simple artificial neural network. (e,f) The manifold $\mathcal{M} \subset \mathbb{R}^{k}$ contains a low-dimensional projection of $\mathbf{r}(t) \in \mathbb{R}^{3N}$ into $\{\psi_i\}_{i=1}^k$ from which molecular configurations $\hat{\mathbf{r}}(t) \in \mathbb{R}^{3N}$ may be approximately reconstructed. We perform the reconstruction by first learning the pairwise distances between all atoms $\mathbf{d} \in \mathbb{R}^{N(N-1)/2}$ using an artificial neural network (panel e) and then using classical multidimensional scaling to deterministically transform this into the reconstructed atomic coordinates $\hat{\mathbf{r}}(t) \in \mathbb{R}^{3N}$ (panel f). All molecular renderings in this work were constructed using VMD \cite{humphrey1996vmd}.}
    \label{fig:pathway}
\end{figure*}

A cartoon schematic of STAR is presented in Fig.~\ref{fig:pathway}. The objective of this work is to train STAR as an accurate and generalizable model to predict molecular configurations from time series data in experimental observables through the four-step pathway $b \rightarrow d \rightarrow c \rightarrow e \rightarrow f$. Each panel in the figure corresponds to a different representation of the molecular system and the arrows between them correspond to mathematical operations to convert one representation to another. Red arrows correspond to unsupervised learning problems, typically nonlinear dimensionality reduction, blue arrows to supervised learning problems requiring learning of a nonlinear function linking the inputs and outputs, and grey arrows to deterministic operations. In order to learn and approximate each of these functions we require training data for which the molecular configurations (Fig.~\ref{fig:pathway}a) corresponding to each point in the time series in the experimental observable (Fig.~\ref{fig:pathway}b) is known. We obtain the former from atomistic MD simulations and the latter by computing the experimentally-measurable observable corresponding to each frame in the trajectory. In this work, we adopt the head-to-tail ($h2t$) distance as an experimental observable that could, in principle, be measured by smFRET \cite{Roy2008rr,Zerze2014mo}. We emphasize that the MD simulation data is only required to train the model and once the model is trained there is no requirement for any additional molecular calculations. The trained model predicts molecular configurations (Fig.~\ref{fig:pathway}f) from novel time series data (Fig.~\ref{fig:pathway}b) that was not present in the training data via the pathway $b \rightarrow d \rightarrow c \rightarrow e \rightarrow f$. In this work, we generate the novel time series data from additional MD simulations, but, in principle, the STAR model trained and calibrated over MD data could be applied to experimental measurements. We detail the additional sophistications that would be necessary to achieve that goal in the final section of the paper. A separate STAR model must be trained for each molecular system.

What is the origin of the multi-step pathway in STAR illustrated in Fig.~\ref{fig:pathway}? Why not predict molecular configurations directly from the $h2t$ time series in a single step $b \rightarrow f$? In this work, we exploit the generic low effective dimensionality of molecular systems along with theoretical guarantees offered by Takens' Delay Embedding Theorem to formulate a succession of simpler and typically lower-dimensional learning problems with firm theoretical underpinnings that we solve using appropriate numerical tools. An additional benefit of this perspective is that the approach also explicitly learns the smFES. The trained STAR model therefore furnishes both a prediction of the molecular structure and its location and thermodynamic stability on the smFES. We now detail each step of the STAR approach presented in Fig.~\ref{fig:pathway}.

\subsubsection{Molecular dynamics training data $(\mathbf{r}(t), v(t))$ }

MD simulations of sufficient duration to comprehensively sample the thermally-relevant configurational space of the molecular system are required to provide the initial training data for STAR. Data that do not sample all relevant conformational states and transitions will result in models overfitted to the training data and a poorly calibrated STAR model that is unable to generalize to new data. For the small macro and biomolecular systems considered in this work, simulations of a few hundred ns to a few $\mu$s were sufficient to provide good sampling and generalizability. The simulation trajectory provides a temporally-ordered sequence of snapshots $\mathbf{r}(t) \in \mathbb{R}^{3N}$ (Fig.~\ref{fig:pathway}a). For each snapshot we compute the instantaneous value of an experimentally-accessible observable to define a 1D time series $v(t) \in \mathbb{R}^1$ (Fig.~\ref{fig:pathway}b). In the present work, we set $v(t)$ to be the head-to-tail molecular distance $h2t(t)$ that is, in principle, measurable by smFRET by conjugating the termini of the molecules with fluorescent probes. These time series can therefore be regarded as idealized synthetic smFRET time traces with no measurement noise and arbitrarily high time resolution. These training data are used to train all steps of the STAR pipeline. 
\subsubsection{Learning the atomistic manifold $\mathcal{M}$}

Interactions between the constituent atoms of the molecular system generically constrain molecular trajectories $\mathbf{r}(t) \in \mathbb{R}^{3N}$ within the 3N-dimensional Cartesian coordinate space to a so-called intrinsic manifold $\mathcal{M} \in \mathbb{R}^{k} \subset \mathbb{R}^{3N}$ of effective dimensionality $k \ll 3N$ \cite{ferguson2010systematic,garcia1992large, amadei1993essential, hegger2007complex, Zhuravlev2009, das2006low,belkin2003laplacian} (Fig.~\ref{fig:pathway}c). We learn this manifold from the simulation trajectory training data using the diffusion maps (dMaps) nonlinear manifold learning technique \cite{coifman2008diffusion, ferguson2010systematic, das2006low, ferguson2011cpl, coifman2005geometric, coifman2006diffusion, lpbeltrami, nadler2006advances,coifman2005geometric,coifman2006diffusion}. Applying dMaps identifies both the dimensionality $k$ of the latent manifold and the CVs $\{\psi_1, \psi_2, \ldots \psi_k\}$ parameterizing it as nonlinear functions of the atomic coordinates \cite{ferguson2010systematic, ferguson2011cpl, Ferguson_2017, Sidky2020}. The dMap CVs are the leading eigenvectors of a discrete random walk constructed over the $3N$-dimensional snapshots comprising the simulation trajectory. The bandwidth $\epsilon$ of the Gaussian kernel used to construct the random walk can be tuned automatically based on the structure of the data \cite{ferguson2010systematic, coifman2008graph} and the dimensionality of the embedding $k$ defined by a gap in the eigenvalue spectrum \cite{ferguson2010systematic, ferguson2011cpl, Wang2017,coifman2006diffusion,coifman2005geometric}. 

Assuming that the distance metric used to measure pairwise similarities between molecular configurations is a good proxy for the kinetic proximity of the configurations over short time scales, these CVs furnish a dynamically meaningful low-dimensional embedding into the slowest relaxing modes of a diffusion process over the data. In particular, Euclidean distances in the embedding correspond to diffusion distances in configurational space that measure the kinetic proximity of any pair of configurations under the action of the random walk \cite{coifman2005geometric,nadler2006advances}. This dynamic interpretability of dMap CVs make them an excellent choice for parameterization of the intrinsic manifold $\mathcal{M}$, but other nonlinear dimensionality reduction or manifold learning techniques such as Isomap \cite{Tenenbaum2000,Saul2006,Li2006,Jwang2011}, locally linear embeddings (LLE) \cite{roweis2000nonlinear,Zhang2006}, local tangent space alignment \cite{Jwang2011} may also be employed.

The only input to the diffusion map algorithm is the definition of a distance metric $k(\mathbf{r}_i,\mathbf{r}_j)$ with which to measure the dissimilarity of pairs of configurations $\left( \mathbf{r}_i, \mathbf{r}_j \right)$ in the trajectory. In order to guarantee the existence of the diffeomorphism, it is critical that this metric be appropriately symmetrized to eliminate any \textit{spatial symmetries} that cannot be distinguished by the choice of experimentally-measurable observable $v(t)$. Our choice of observable $v(t) = h2t(t)$ is invariant under translation, rotation, mirror inversion, and -- for chemically symmetric molecules (e.g., simple polymer chains) -- head-tail inversion of the molecular configuration $\mathbf{r} \in \mathbb{R}^{3N}$. Any representation of the system based on $v(t)$ sacrifices the ability to distinguish changes in the system state under any of these transformations. It is possible to recover some or all of these symmetries under different choices for $v(t)$ or by multiplexed simultaneous measurements $\mathbf{v}(t) = \{v_1(t), v_2(t), \ldots\}$. As is standard practice in measuring molecular similarity, we select $k(\mathbf{r}_i,\mathbf{r}_j)$ to be the rotationally and translationally aligned root mean squared deviation (RMSD), which naturally mods out the rototranslational symmetry using the Kabsh algorithm \cite{kabsch1976solution}, and we eliminate the discrete symmetries by minimizing under mirror inversion,
\begin{align}
k(\mathbf{r}_i,\mathbf{r}_j) &= \underset{\textrm{mirror inversion}}{min} \; RMSD(\mathbf{r}_i,\mathbf{r}_j).
\end{align}
or both mirror and head-tail inversion for chemically symmetric molecules \cite{Ferg18}, 
\begin{align}
k(\mathbf{r}_i,\mathbf{r}_j) &= \underset{\textrm{head-tail inversion}}{min} \; \underset{\textrm{mirror inversion}}{min} \; RMSD(\mathbf{r}_i,\mathbf{r}_j).
\end{align}
Failure to eliminate all such spatial symmetries induced by the choice of observable $v(t)$ violates Takens' Theorem and can cause the manifolds $\mathcal{M}$ and $\mathcal{M}^\prime$ to no longer be diffeomorphisms related by a well defined and learnable transformation.

The dMap CVs $\{\psi_i\}_{i=1}^k$ define a data-driven parameterization of the $k$-dimensional manifold $\mathcal{M}$. Projection of the full simulation trajectory into these CVs defines an empirical probability distribution $P(\psi_1, \psi_2, \ldots \psi_k)$. An estimate of the smFES mapping out the free energy $F$ over $\mathcal{M}$ is obtained via the statistical mechanical relationship $\beta F(\psi_1, \psi_2, \ldots \psi_k) = -\ln P(\psi_1, \psi_2, \ldots \psi_k) + C$, where $C$ is an arbitrary additive constant and the free energy is de-dimensionalized by the reciprocal temperature $\beta = 1/k_B T$. New molecular configurations $\mathbf{r}_\textrm{new}$ not contained within the data used to construct $\mathcal{M}$ may be projected onto the manifold using the Nystr\"om extension \cite{sonday2009coarse,laing2007coarse,long2019landmark}.

\subsubsection{Learning the Takens' manifold $\mathcal{M}^\prime$}

The previous section detailed a procedure to estimate the intrinsic manifold $\mathcal{M}$ and the smFES it supports by applying manifold learning to a simulation trajectory of the atomistic molecular configuration $\mathbf{r}(t) \in \mathbb{R}^{3N}$. Takens' Delay Embedding Theorem \cite{Takens,sauer1991embedology,packard1980,Broomhead1986,Cao1998,stark2003delay,complex1,holger,PMID:25733874} provides a means to reconstruct an image of the intrinsic manifold $\mathcal{M}^\prime$ (Fig.~\ref{fig:pathway}d) by applying similar operations to delay embeddings of a time series in a single coarse-grained observable $v(t) \in \mathbb{R}^1$ (Fig.~\ref{fig:pathway}b). Takens' Theorem asserts that (i) a $(d \geq (2k+1))$-dimensional delay embedding $\mathbf{y}(t) = [v(t), v(t-\tau), v(t-2\tau), \ldots , v(t-(d-1)\tau)] \in \mathbb{R}^d$ in a generic observable $v(t)$ constructed at a delay time $\tau$ uniquely specifies the instantaneous state of the system, (ii) the dynamical evolution of $\mathbf{y}(t)$ is $C^1$-equivalent (i.e., identical under a smooth and continuous mapping) to that of $\mathbf{r}(t)$, and (iii) the evolution of $\mathbf{y}(t)$ lies on a manifold $\mathcal{M}^\prime$ that is related to $\mathcal{M}$ by a diffeomorphism (i.e., smooth, invertible, and bijective transformation) $\Theta : \mathcal{M}^\prime \rightarrow \mathcal{M}$ with inverse $\Theta^{-1} : \mathcal{M} \rightarrow \mathcal{M}^\prime$ \cite{Ferg18,Takens,sauer1991embedology,packard1980,Broomhead1986,Cao1998,stark2003delay,complex1,holger,Zerze2014mo}. The diffeomorphism is \textit{a priori} unknown but is guaranteed only to stretch and squash the manifold and not tear it or stitch it together in new ways. As such, $\mathcal{M}^\prime$ is a topologically identical image of $\mathcal{M}$ that preserves its continuity and connectivity \cite{Takens,sauer1991embedology, packard1980, Broomhead1986, complex1, holger}. 

Takens' Theorem holds for any generic observable of the system $v$ that does not contain any spurious symmetries that are not present in the system itself (i.e., the system is invariant in the observable under particular symmetries), for any delay time $\tau$ that does not introduce temporal aliasing (i.e., is a multiple of a period of the dynamical motion), and for any delay embedding dimensionality $d$ greater than twice the intrinsic dimension of the system $k$ \cite{Takens,sauer1991embedology,holger,letellier1998non,cross2010differential,letellier1996topological,letellier1998non,Ferg16,Ferg18}. In practice, better results are obtained for observables $v$ that are strong functions of all system degrees of freedom and which respond sensitively to the important dynamical motions of the system \cite{Ferg18}, for delay times $\tau$ estimated as the first minimum of the autocorrelation or mutual information of $v(t)$ \cite{Fraser86,MI}, and for delay dimensionalities $d$ estimated using the E$_1$ method of Cao \cite{Cao97,VillaniB,kennel1992determining}. Takens' Theorem still holds when applied to observables $v$ that do contain symmetries not present in the system, but the system may only be reconstructed up to those symmetric operations (\textit{vide supra}), and also to observables of subsystems and under stochastic or deterministic forcing \cite{stark1999delay,stark2003delay}. The latter two generalizations are relevant to the present work because we adopt the head-to-tail molecular distance as our observable $v(t) = h2t(t)$, which is both an observation of the solute subsystem within the full solute-solvent system and is subject to dynamical coupling with solvent and deterministic or stochastic forcing by any attached thermostats, barostats, or other external constraints. Finally, Takens' Theorem may also be applied to multiplexed measurements of several simultaneous observables $\mathbf{v}(t) = \{v_1(t), v_2(t), \ldots\}$ such as multi-channel smFRET measuring multiple intramolecular distances \cite{Roy2008rr,Cao1998}.

The manifold $\mathcal{M}^\prime \subset \mathbb{R}^{k} \subset \mathbb{R}^{d}$ is estimated by applying dMaps to the delay embedding $\mathbf{y}(t) = [v(t), v(t-\tau), v(t-2\tau), \ldots , v(t-(d-1)\tau)]$ parameterized by the $k$ CVs $\{\psi_1^\prime, \psi_2^\prime, \ldots \psi_k^\prime\}$. We apply dMaps under a distance metric $k^\prime(\mathbf{y}_i,\mathbf{y}_j)$ measuring the dissimilarity of pairs of delay vectors that we select to be a simple Euclidean distance metric. This metric requires modification due to the introduction of a spurious symmetry into the system by the temporal ordering of $v(t)$ within the $\mathbf{y}$ vectors induced by the Takens' delay embedding. This spurious symmetry can be understood by considering a hypothetical microstate transition of the system $\mathbf{r}_A(t-\tau) \rightarrow \mathbf{r}_B(t) \rightarrow \mathbf{r}_C(t+\tau)$ and its reverse $\mathbf{r}_C(t-\tau) \rightarrow \mathbf{r}_B(t) \rightarrow \mathbf{r}_A(t+\tau)$, with associated $d$=3-dimensional delay vectors $\mathbf{y}_\mathrm{fwd} = [v_A, v_B, v_C]$ and $\mathbf{y}_\mathrm{bkwd} = [v_C, v_B, v_A]$. Adopting a convention that defines a mapping between system microstates and delay vectors based on the central microstate, both of these delay vectors are associated with microstate $\mathbf{r}_B$ but the delay vectors themselves are not identical. By observing the past and future of $\mathbf{r}_B$ we can distinguish whether it was occupied as part of a forward or reverse transition, and the ``forward'' $\mathbf{r}_B$ and ``backward'' $\mathbf{r}_B$ appear in the Takens' delay embedding as distinct identifiable states. For equilibrium systems obeying detailed balance, the forward and backward transitions between any pair of microstates are equally probable and we should not be able to identify whether occupancy of a particular microstate $\mathbf{r}_B$ resulted from the forward or reverse transition in any such pair. To prevent the occurrence of this symmetry breaking in $\mathcal{M}^\prime$ that remains unbroken in $\mathcal{M}$ we must symmetrize $k^\prime(\mathbf{y}_i,\mathbf{y}_j)$ to eliminate this \textit{temporal symmetry}. This assures that Takens' Theorem is not violated and the manifolds $\mathcal{M}$ and $\mathcal{M}^\prime$ remain diffeomorphic. We have previously demonstrated the importance of eliminating this symmetry in applications of Takens' Theorem to equilibrium molecular systems \cite{Ferg16,Ferg18}. As such, we minimize the distance metric under time reversal, which is equivalent to minimizing under inversion of one of the delay vectors,
\begin{align}
k^\prime(\mathbf{y}_i,\mathbf{y}_j) &= min \left[ \norm{\mathbf{y}_i - \mathbf{y}_j}_2, \norm{\mathbf{y}_i - flip\left(\mathbf{y}_j\right)}_2 \right]
\end{align}

The dMap CVs $\{\psi_i^\prime\}_{i=1}^k$ define a data-driven parameterization of the $k$-dimensional manifold $\mathcal{M}^\prime$ over which we construct the smFES $\beta F^\prime(\psi_1^\prime, \psi_2^\prime, \ldots \psi_k^\prime) = -\ln P^\prime(\psi_1^\prime, \psi_2^\prime, \ldots \psi_k^\prime) + C^\prime$ by compiling the empirical probability distribution $P^\prime(\psi_1^\prime, \psi_2^\prime, \ldots \psi_k^\prime)$ of the delay vectors projected onto the manifold. New delay vectrors $\mathbf{y}_\textrm{new}$ not contained within the data used in the construction of $\mathcal{M}^\prime$ may be projected onto the manifold using the Nystr\"om extension \cite{sonday2009coarse,laing2007coarse,long2019landmark}. We exploit this out-of-sample extension when applying the trained STAR model to new data that was not present during training.

\subsubsection{Learning the diffeomorphism from $\mathcal{M}^\prime$ to $\mathcal{M}$} 

Takens' Theorem guarantees the smooth manifolds $\mathcal{M}$ and $\mathcal{M}^\prime$ are related by a diffeomorphism $\Theta : \mathcal{M}^\prime \rightarrow \mathcal{M}$ \cite{Ferg18,Takens,sauer1991embedology,packard1980,Broomhead1986,Cao1998,stark2003delay,complex1,holger,Zerze2014mo}. The theoretical guarantees on the existence of this mapping and its low dimensional nature are a valuable advantage of the multi-step STAR pathway that would be lost by formulating a direct reconstruction of the molecular configurations from the univariate time series. We adopt a convention associating each delay vector $\mathbf{y}(t) = [v(t), v(t-\tau), v(t-2\tau), \ldots, v(t-(d-1)\tau)]$ with the configurational microstate corresponding its central element $\mathbf{r}(t-((d-1)/2)\tau)$. We assert that $d$ be odd in order to make this association unambiguous. This mapping means that the configurations in the leading $t < (d-1)\tau/2$ and trailing $t > (d-1)\tau/2$ periods of the molecular simulation trajectory are not associated with any delay vector and are eliminated from all analyses.

Having defined the associations between the projections of the delay vectors $\mathbf{y}(t)$ on $\mathcal{M}^\prime$ represented as $\{\psi_i^\prime(t)\}_{i=1}^k$ and the projections of the molecular configurations $\mathbf{r}(t)$ on $\mathcal{M}$ represented as $\{\psi_i(t)\}_{i=1}^k$, we define a supervised learning problem between pairs of data points $\left( \{\psi_i^\prime(t)\}_{i=1}^k, \{\psi_i(t)\}_{i=1}^k \right)$ to perform data-driven estimation of the $k$-dimensional to $k$-dimensional diffeomorphism $\Theta : \mathcal{M}^\prime \rightarrow \mathcal{M}$ (Fig.~\ref{fig:pathway}d $\rightarrow$ c). There are many ways to learn and approximate this function, including k-nearest neighbors \cite{Cover1967}, kernel methods \cite{Scholkopf2002}, or local Jacobians \cite{Ferg16,Ferg18,Principlesriemannian}. In this work we employ simple fully-connected feedforward artificial neural networks (ANN) as an easy to train and flexible function approximator \cite{Hassoun1996}. We typically find that networks comprising 4-8 hidden layers each containing $\sim$10$k$ neurons are adequate to furnish high accuracy mappings.

\subsubsection{Learning the reconstruction $\hat{\mathbf{r}}(t)$} 

The final step is to learn approximate reconstructions of the molecular configurations $\mathbf{r}(t) \in \mathbb{R}^{3N}$ from their projections $\{\psi_i(t)\}_{i=1}^k \in \mathbb{R}^k$ onto the intrinsic manifold $\mathcal{M}$. If the effective dimensionality of the simulation trajectory is less than or equal to $k$ and the dMap CVs have been properly learned from the simulation trajectory, then the $k$-dimensional subspace spanned by $\{\psi_i(t)\}_{i=1}^k$ is expected to preserve the important configurational variance in $\mathbf{r}(t)$  \cite{ferguson2010systematic,ferguson2011cpl}. As such, the location on the intrinsic manifold $\mathcal{M}$ should contain sufficient information to approximately reconstruct the configurational state of the system $\hat{\mathbf{r}}(t) \in \mathbb{R}^{3N}$ up to any symmetries that have been modded out in its construction (\textit{vide supra}). In this sense, the  process $\mathbf{r}(t) \rightarrow \{\psi_i(t)\}_{i=1}^k \rightarrow \hat{\mathbf{r}}(t)$ may be viewed as the concatenation of a low-dimensional encoder furnished by diffusion maps and a decoder to be learned from the data. Again we have the one-to-one mapping of data pairs $\left( \{\psi_i(t)\}_{i=1}^k, \mathbf{r}(t) \right)$ and can formulate and solve this as a supervised learning problem.

We split this learning problem into two steps and instead of learning to predict the Cartesian coordinates of the atoms $\mathbf{r}(t)$ directly from $\{\psi_i(t)\}_{i=1}^k$, we first learn the ordered vector of pairwise distances between all atoms $\mathbf{d}(t) \in \mathbb{R}^{N(N-1)/2}$. Rigid graph theory asserts that specification of all $N \choose 2$ pairwise distances defines the absolute coordinates of the $N$ points up to translation, rotation, and mirror inversion \cite{Euclid,Singer2008}. In practice, calculation of the coordinates from the pairwise distances is easily accomplished using classical multidimensional scaling (cMDS) to form the Gram matrix and compute its eigendecomposition \cite{Dokmanic_2015,CRIPPEN1978449}. In the present application, the translational, rotational, and mirror inversions (and head-tail inversion, where applicable) have all been modded out of the construction of $\mathcal{M}$ and so there is no (additional) information loss in formulating the supervised learning problem as first approximating the functional mapping $\{\psi_i(t)\}_{i=1}^k \rightarrow \mathbf{d}(t)$ (Fig.~\ref{fig:pathway}c $\rightarrow$ e) and then making the deterministic transformation $\mathbf{d}(t) \rightarrow \hat{\mathbf{r}}$ using cMDS (Fig.~\ref{fig:pathway}e $\rightarrow$ f). Unlike the one-step formulation of the reconstruction $\{\psi_i(t)\}_{i=1}^k \rightarrow \hat{\mathbf{r}}$, the two-step formulation eliminates the need to perform any alignment of the molecular configurations with respect to rotation, translation, and mirror inversion since these symmetries are naturally modded out within the pairwise distance matrix. The omission of these alignment operations, either mutually or to some fixed reference structure, carries advantages in avoiding the introduction of noise and approximations into the fitting problem. 

We use simple ANNs to learn $\{\psi_i(t)\}_{i=1}^k \rightarrow \mathbf{d}(t)$ from the training data. Typically we find that networks comprising 4-8 hidden layers each containing on the order of up to $\sim$1000$k$ neurons are adequate to furnish high accuracy mappings. The molecular configurations $\hat{\mathbf{r}}(t) \in \mathbb{R}^{3N}$ are reconstructed only up to translational, rotational, and mirror inversion symmetries (and head-tail inversion, for chemically symmetric molecules), and so comparisons of the reconstruction accuracy between $\left( \hat{\mathbf{r}}, \mathbf{r} \right)$ pairs must be performed by mutual alignment under these transformations. The alignment problem corresponds to an orthogonal Procrustes problem \cite{Schnemann1966AGS} that we efficiently solve using the Kabsch algorithm \cite{kabsch1976solution}.

\subsubsection{Deploying the trained STAR model} 

The STAR pipeline trained for a particular molecular system may then be used to predict molecular reconstructions $\hat{\mathbf{r}}(t) \in \mathbb{R}^{3N}$ from new time series data $v(t) \in \mathbb{R}^1$ through the four-step pathway $b \rightarrow d \rightarrow c \rightarrow e \rightarrow f$ (Fig.~\ref{fig:pathway}b-f): Takens' delay vectors $\mathbf{y}$ are constructed from the time series $v(t)$, projected onto the Takens' manifold $\mathcal{M}^\prime$ using the Nystr\"om extension \cite{sonday2009coarse,laing2007coarse,long2019landmark}, mapped onto the atomistic manifold $\mathcal{M}$ using the trained ANN, converted into the atomistic pairwise distances matrix $\mathbf{d}$ using another trained ANN, and then finally transformed into the reconstructed molecular configurations $\hat{\mathbf{r}}$ using cMDS. Provided the training data were sufficiently rich to span the thermally-relevant metastable states and transitions in the system and the supervised learning problems are not overfitted, then the trained STAR model should be able to generalize to new time series data from the system that were not included in model training. Importantly, deployment of the trained model does not require any additional molecular simulation data during the deployment phase (Fig.~\ref{fig:pathway}a). 

We validate the trained model by extracting time traces of the molecular heat-to-tail distance $v(t) = h2t(t)$ from independent MD simulation trajectories that were not present in the training data and testing the capability of the trained STAR model to reconstruct the molecular configurations $\mathbf{r}(t)$ in the trajectory. A well-trained model should be able to predict molecular configurations from novel testing data with similar accuracies to that in the training data. In applications to real experimental smFRET data, the true molecular configurations would not be available to make this comparison and the model predictions would have to be validated by indirect means. In all cases we seek only to reconstruct the molecular configuration of the solute and do not attempt to predict the coordinates of the water solvent or any counter ions. In principle, this information is -- again subject to the relevant symmetries -- present within the Takens' delay embedding, but is much more challenging to recover due to the permutational fungibility of the water molecules and the relatively weaker influence of solvent coordinates on our choice of solute-centric observable.

\subsection{Molecular Dynamics Simulations} \label{subsec:MD}

\subsubsection{\ce{C_{24}H_{50}}} \label{subsubsec:meth:C24}

MD simulations of \ce{C_{24}H_{50}} were conducted using the Gromacs 4.6 simulation suite \cite{Gromacs} using the TraPPE potential \cite{martin1998transferable} that models each \ce{CH_3} and \ce{CH_2} group as a united atom and the all-atom SPC model of water \cite{water}. Chain topologies were constructed using the PRODRG2 server \cite{PRODRG}. Lennard Jones interactions were smoothly set to zero at a cutoff of 1.4 nm and Lorentz-Berthelot combining rules used to determine dispersion interactions between unlike atoms \cite{allen1989computer}. Electrostatic interactions were treated using particle mesh Ewald \cite{essmann1995smooth} with a real-space cutoff of 1.4 nm and a 0.12 nm reciprocal-space grid spacing that were optimized during runtime. Simulations were conducted in a 5$\times$5$\times$5 nm$^3$ cubic box with periodic boundary conditions that was sufficiently large to prevent self-interactions of the chain through the periodic walls even in its fully-extended configuration. Initial system configurations were generated by placing an initially elongated chain into an empty box and solvating with 4117 water molecules to a density of 1.0 g/cm$^3$. High energy overlaps were eliminated by steepest descent energy minimization to a force threshold of 2000 kJ/mol.nm. Initial particle velocities were sampled from a Maxwell-Boltzmann distribution at 298 K. Simulations were performed in the NPT ensemble at 298 K and 1 bar using a Nos\'e-Hoover thermostat \cite{nose1984unified} and an isotropic Parrinello-Rahman barostat \cite{parrinello1981polymorphic}. Equations of motion were integrated using the leap-frog algorithm \cite{hockney2010computer} with a 2 fs time step. LINCS constraints were used to fix bond lengths to their equilibrium distances for computational efficiency and as required by the TraPPE and SPC potentials \cite{nadler2006advances}. Systems were equilibrated for 1 ns before conducting a 100 ns production run during which system configurations were saved every 0.2 ps. Head-to-tail distances were computed for each frame of the 500,000 frame simulation trajectory to furnish the univariate time series $v(t) = h2t(t)$ in addition to the atomistic simulation trajectory $\mathbf{r}(t)$. The first 20 ns (100,000 frames) of these trajectories provided the $\left( \mathbf{r}(t), v(t) = h2t(t) \right)$ training data used to train the STAR model, and the remaining 80 ns (400,000 frames) reserved for testing. Input files for these simulations are provided in the \blauw{Supplementary Material}.

\subsubsection{Chignolin} \label{subsubsec:meth:Chig}

MD simulations of the 10 residue (166 atom) engineered mini-protein Chignolin (GYDPETGTWG, PDB ID: 1UAO) \cite{Honda2004} were performed by D.E.~Shaw Research using the Desmond simulation suite \cite{4090217} on the Anton supercomputer \cite{Shaw341} and reported in Ref.~\cite{deshaw}.  The peptide was modeled using the the CHARMM22$\*$ force field \cite{PMID:21539772} and solvated by $\sim$1900 water molecules \cite{jorgensen1983comparison} in a 4$\times$4$\times$4 nm$^3$ cubic box with periodic boundary conditions. Two Na$^+$ ions were added to maintain charge neutrality. Lennard-Jones interactions were treated with a 0.95 nm cutoff and electrostatic interactions treated by Gaussian Split Ewald \cite{doi:10.1063/1.1839571} employing a 0.95 nm real-space cutoff and a $32 \times 32 \times 32$ cubic grid. Equations of motion were integrated with a 2.5 fs time step. Equlibration runs were conducted in the NPT ensemble at 340 K and maintained by a Nos\'e-Hoover thermostat \cite{nose1984unified,Hoover}. The 106 $\mu$s production run was conducted in the NVT ensemble employing a Nos\'e-Hoover thermostat \cite{nose1984unified,Hoover} and system configurations harvested every 200 ps. The first 5 $\mu$s (25,000 frames) of these trajectories were used to train the STAR model, and the next 15 $\mu$s (75,000 frames) employed for testing.

\section{Results and Discussion} \label{sec:results}

We demonstrate and validate STAR in applications to molecular dynamics simulations of a \ce{C_{24}H_{50}} polyethylene chain and the $\beta$-hairpin engineered mini-protein Chignolin. Simulation data constituting the training and testing data are collected as described in Section \ref{subsec:MD}. The STAR pipeline was trained and deployed as described in Section \ref{subsec:STAR} with details specific to each system provided below. In each case we achieve RMSD reconstruction accuracies on the hold-out test data of better than 0.2 nm.

\subsection{\ce{C_{24}H_{50}}} \label{subsec:res:C24}

\subsubsection{STAR Training}

The training portion of the \ce{C_{24}H_{50}} simulation trajectory comprised 100,000 frames saved at 0.2 ps intervals recording the Cartesian coordinates $\mathbf{r}(t) \in \mathbb{R}^{72}$ of the $N$ = 24 united atoms of the polymer chain (cf.\ Fig.~\ref{fig:pathway}a). In order to run dMaps into local memory, the simulation trajectory was subsampled with a stride of 20 and the excluded frames projected into the manifold \textit{post hoc} using the the Nystr\"om extension \cite{sonday2009coarse,laing2007coarse,long2019landmark}. Application of spatially-symmetrized dMaps to these data employing a kernel bandwidth of $\epsilon$ = $\exp(-3)$ nm exposed a $k$=2-dimensional intrinsic manifold $\mathcal{M} \subset \mathbb{R}^2$ spanned by nonlinear collective variables of the united atom coordinates $\{ \psi_1, \psi_2 \}$ and supporting the smFES $\beta F(\psi_1, \psi_2)$ (cf.\ Fig.~\ref{fig:pathway}c). 

A scalar time series in the head-to-tail distance between the two terminal united atoms $v(t) = h2t(t) \in \mathbb{R}^1$ was computed from the MD training trajectory as a synthetic and idealized smFRET time series over the 100,000 frames separated by intervals of 0.2 ps (cf.\ Fig.~\ref{fig:pathway}b).  Takens' delay vectors $\mathbf{y}(t) = [h2t(t), h2t(t-\tau), h2t(t-2\tau), \ldots , h2t(t-(d-1)\tau)]$ were constructed at a delay time of $\tau$ = 30 ps (150 time steps) computed as the first minimum in the mutual information of $h2t(t)$ \cite{Fraser86} and delay dimensionality $d$ = 5 computed using the E$_1$ method of Cao \cite{Cao97,VillaniB,kennel1992determining}. This resulted in the construction of 99,400 delay vectors, each associated with a frame in the MD trajectory containing the molecular structure from which the central element in the delay vector was computed. The initial and terminal $(d-1)\tau/2$ = 60 ps (300 frames) of the MD trajectory that were unassociated with any delay vector were dropped from further analyses. Application of temporally-symmetrized dMaps to the delay embedding trajectory, subsampled with a stride of 20 in order to fit into local memory and employing a kernel bandwidth of $\epsilon$ = 1 nm, defined a $k$ = 2-dimensional manifold $\mathcal{M}^\prime \subset \mathbb{R}^2$ spanned by $\{ \psi_1^\prime, \psi_2^\prime \}$. The frames strided over in the application of dMaps were projected into $\mathcal{M}^\prime$ using the Nystr\"om extension and used to estimate the smFES $\beta F^\prime(\psi_1^\prime, \psi_2^\prime)$ (cf.\ Fig.~\ref{fig:pathway}d).

The diffeomorphism $\Theta : \mathcal{M}^\prime \subset \mathbb{R}^2 \rightarrow \mathcal{M} \subset \mathbb{R}^2$ linking the two manifolds was learned using a simple 2-25-25-25-25-2 fully-connected feedforward ANN comprising four hidden layers of 25 neurons (cf.\ Fig.~\ref{fig:pathway}d $\rightarrow$ c). The ANN employed tanh activations and was trained using Adam \cite{kingma2014adam} with a batch size of 500 and a learning rate of 1$\times$10$^{-4}$. Training was terminated after 250 epochs upon plateauing of the validation error. 

The function linking the projection of each frame in the MD trajectory into the intrinsic manifold $\{ \psi_1(t), \psi_2(t) \}$ to the $N(N-1)/2$ = 276-dimensional united atom pairwise distance vectors $\mathbf{d}(t) \in \mathbb{R}^{276}$ corresponding to that configuration was learned using a 2-4-187-370-552-276 fully-connected feedforward ANN employing tanh activations and trained using Adam \cite{kingma2014adam} with a batch size of 500 and a learning rate of 1$\times$10$^{-3}$ over 150 epochs (cf.\ Fig.~\ref{fig:pathway}c $\rightarrow$ e). Molecular reconstructions of the united atom coordinates $\hat{\mathbf{r}}(t) \in \mathbb{R}^{72}$ were computed deterministically from each pairwise distance vector using cMDS (cf.\ Fig.~\ref{fig:pathway}e $\rightarrow$ f). The reconstruction accuracy of the trained STAR pipeline (cf.\ Fig.~\ref{fig:pathway}b $\rightarrow$ d $\rightarrow$ c $\rightarrow$ e $\rightarrow$ f) is assessed by computing the RMSD of the predicted $\hat{\mathbf{r}}(t) \in \mathbb{R}^{72}$ and true $\mathbf{r}(t) \in \mathbb{R}^{72}$ configurations under translational, rotational, mirror, and head-tail alignment by the Kabsch algorithm \cite{kabsch1976solution}.

\subsubsection{STAR Deployment}

\begin{figure*}[t]
    \centering
    \includegraphics[width=0.9\textwidth]{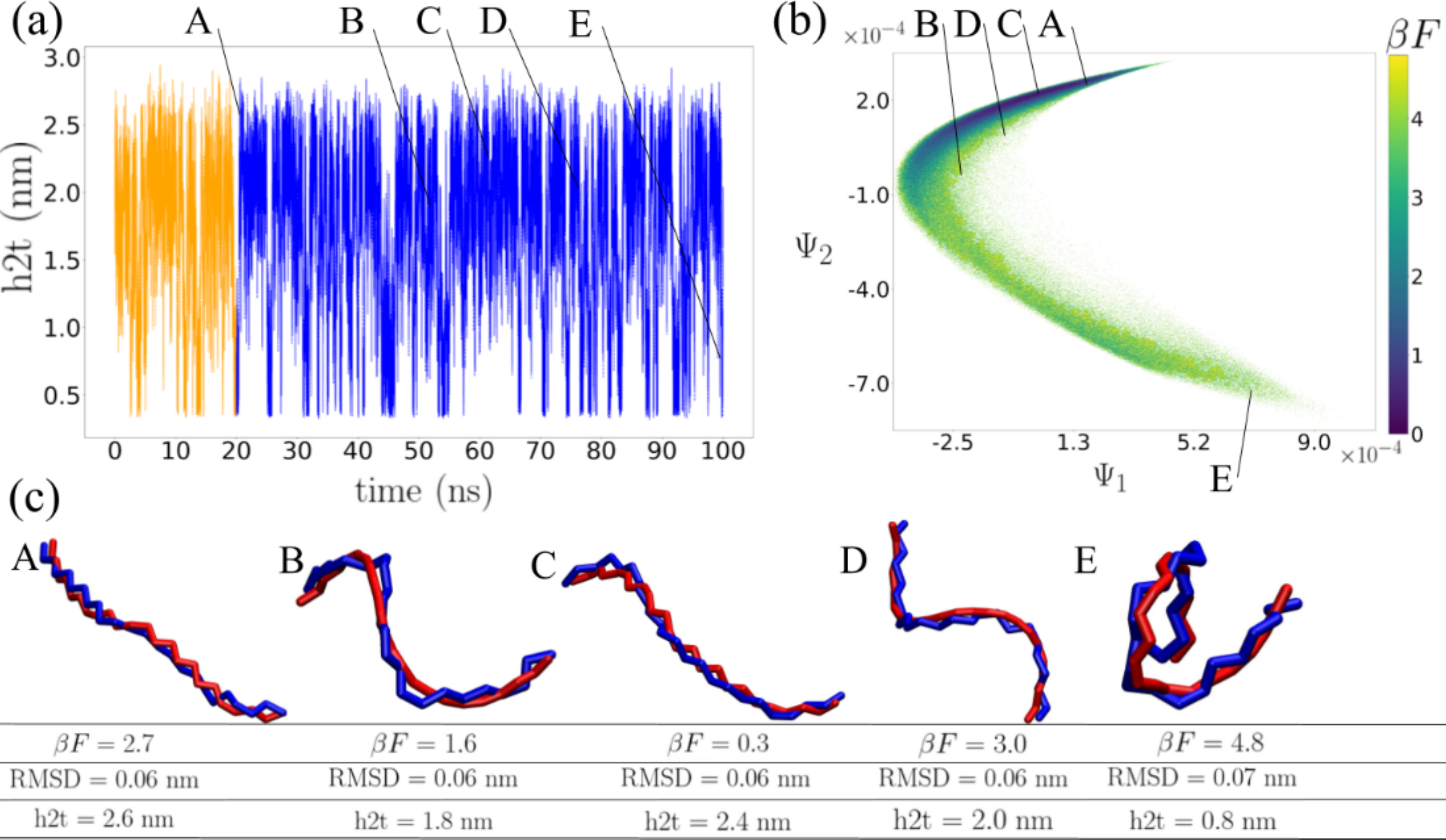}
    \caption{Application of STAR to \ce{C_{24}H_{50}} polymer chain. (a) Synthetic idealized smFRET time trace of the head-to-tail distance $h2t(t)$ between the terminal united atoms computed over a 100 ns MD trajectory with frames saved every 0.2 ps. The first 20 ns are used for training (orange) and the remaining 80 ns for testing (blue). The molecule undergoes hundreds of folding and unfolding events over the course of the simulation. (b) The intrinsic manifold $\mathcal{M}$ spanned by the dMap CVs $\{ \psi_1(t), \psi_2(t) \}$ and supporting the smFES $\beta F(\psi_1, \psi_2)$. The Gibbs free energy $F$ is dedimensionalized by the inverse temperature $\beta = 1/k_B T$. The arbitrary zero of $F$ is specified to lie at the global free energy minimum. (c) Molecular reconstructions using the trained STAR pipeline of five representative points A-E in the $h2t(t)$ time series spanning the test set. Reconstructions $\hat{\mathbf{r}}$ (red) are superposed on the corresponding true configurations $\mathbf{r}$ (blue) extracted directly from the MD simulation. The head-to-tail distance of the true configuration and the RMSD under translational, rotational, mirror, and head-tail alignment between the true and reconstructed configurations are reported under each image.  The STAR prediction of the location of each point on the smFES is illustrated in panel b and the corresponding dimensionless free energy $\beta F$ reported under each image.} \label{fig:C24}
\end{figure*}

We illustrate the application of the trained \ce{C_{24}H_{50}} STAR model in Fig.~\ref{fig:C24}. The trained STAR pipeline enables us to associate each value of $h2t(t)$ in the synthetic smFRET time trace with a molecular reconstruction and also a location and stability on the smFES supported by the low-dimensional intrinsic manifold~\cite{ferguson2010systematic,Ferg16}. The RMSD reconstruction accuracy of the trained pipeline applied to the 20 ns training trajectory is RMSD$_\mathrm{train}$ = 8.4 $\times$10$^{-2}$ nm. In an application to the remaining 80 ns testing trajectory that was not part of the training ensemble, the reconstruction accuracy degrades only slightly to RMSD$_\mathrm{test}$ = 8.6 $\times$10$^{-2}$ nm, indicating that the STAR model is well trained and has good generalizability to novel data. We illustrate for five selected points A-E in the $h2t(t)$ time trace (Fig.~\ref{fig:C24}a) their projection onto the smFES $\beta F(\psi_1, \psi_2)$ (Fig.~\ref{fig:C24}b) and their reconstructed $\hat{\mathbf{r}}(t) \in \mathbb{R}^{72}$ and true $\mathbf{r}(t) \in \mathbb{R}^{72}$ molecular structures (Fig.~\ref{fig:C24}c). \blauw{Movie S1} in the \blauw{Supplementary Material} presents an animation of molecular reconstructions and smFES projections for all data in the time series in Fig.~\ref{fig:C24}a.

Configuration A corresponds to an elongated chain configurations with the preponderance of the backbone dihedrals in the trans state. Configuration C corresponds to a similarly elongated conformation but with a small number of gauche defects that lead to a small degree of curvature in the contour of the chain. The elevated conformational entropy associated with this small degree of curvature allows Configuration C to reside within 0.3 $k_B T$ of the global free energy minimum at $\beta F$ = 0, while Configuration A lies slightly higher at $\beta F$ = 2.7. Configurations B and D correspond to partially collapsed twisted structures $\sim$3 $k_B T$ less stable than the global free energy minimum that lie along the transition pathway between the global minimum at $(\psi_1 \sim 1.2, \psi_2 \sim 2.0)$ containing elongated chains and the weak local minimum at at $(\psi_1 \sim 3.2, \psi_2 \sim -6.0)$ containing hydrophobically collapsed coiled chains. Configuration E is a metastable hydrophobically collapsed coil that lies slightly outside the local minimum at $\beta F$ = 4.8. We note that Configurations B ($h2t$ = 1.8nm) and D ($h2t$ = 2.0 nm) possess similar values of the $h2t$ distance but correspond to very distinct molecular configurations residing at different locations on the smFES that are both accurately reconstructed by STAR. This illustrates the value of using Takens' Theorem to reconstruct molecular configurations that cannot be distinguished from the instantaneous value of the observable alone.

\subsection{Chignolin} \label{subsec:res:Chig}

\subsubsection{STAR Training}

The training portion of the Chignolin trajectory comprised 25,000 frames saved at 200 ps intervals recording the Cartesian coordinates $\mathbf{r}(t) \in \mathbb{R}^{279}$ of the $N$ = 93 heavy atoms of the protein. A $k$ = 2-dimensional intrinsic manifold $\mathcal{M} \subset \mathbb{R}^2$ spanned by $\{ \psi_1, \psi_2 \}$ was constructed by applying dMaps with a kernel bandwidth of $\epsilon$ = $\exp(-3)$ nm to a subsampling of thsi trajectory with a stride of 2. The excluded frames were projected into the manifold using the the Nystr\"om extension. Taken's delay embeddings were constructed from the synthetic smFRET time series recording the distance between the terminal heavy atoms at a delay time $\tau$ = 200 ps (1 time step) \cite{Fraser86} and delay dimensionality $d$ = 11 \cite{Cao97,VillaniB,Cao97,kennel1992determining}. This results in the construction of 24,990 delay vectors. Applying temporally-symmetrized dMaps to the delay embedding trajectory, sub-sampled with a stride of 2 in order to fit into local memory, $\mathbf{y}(t) \in \mathbb{R}^{11}$ with a kernel bandwidth of $\epsilon$ = 1 nm defined a $k$ = 2-dimensional manifold $\mathcal{M}^\prime \subset \mathbb{R}^2$. Again, the excluded frames were projected into the manifold using the the Nystr\"om extension. The diffeomorphism $\Theta$ mapping $\mathcal{M}^\prime$ to $\mathcal{M}$ was approximated by a 2-25-25-25-25-2 ANN trained using Adam \cite{kingma2014adam} with a batch size of 500 and learning rate of 1$\times$10$^{-4}$ over 250 epochs. The function mapping locations on $\mathcal{M}$ to the $N(N-1)/2$ = 4278-dimensional heavy atom pairwise distances vectors $\mathbf{d}(t) \in \mathbb{R}^{4278}$ was learned and approximated by a 2-4-2855-5706-8556-4278 ANN using Adam \cite{kingma2014adam} with a batch size of 500 and learning rate of 1$\times$10$^{-5}$ over 150 epochs. Predictions of heavy atom molecular configurations were computed deterministically from the pairwise distance vectors using cMDS.

\subsubsection{STAR Deployment}

\begin{figure*}[t]
    \centering
    \includegraphics[width=0.9\textwidth]{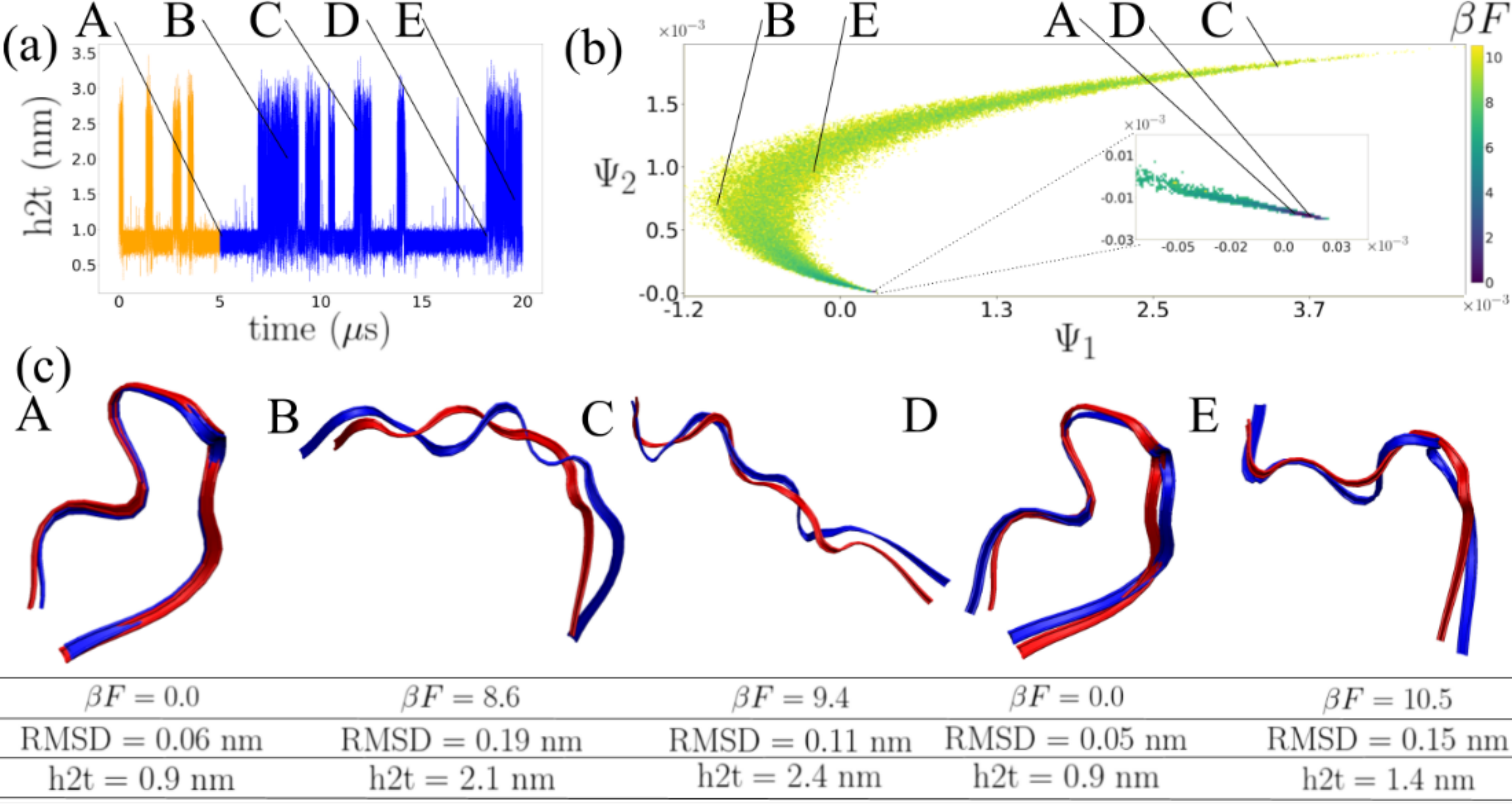}
    \caption{ Application of STAR to the 10 residue engineered mini-protein Chignolin. (a) Synthetic idealized smFRET time trace of the head-to-tail distance $h2t(t)$ between the terminal heavy atoms computed over a 20 $\mu$s MD trajectory with frames saved every 200 ps. The first 5 $\mu$s are used for training (orange) and the remaining 15 $\mu$s for testing (blue). The molecule undergoes dozens of folding and unfolding events over the course of the trajectory. (b) The smFES $\beta F(\psi_1, \psi_2)$ supported by the intrinsic manifold $\mathcal{M}$ has its arbitrary zero of $F$ specified to lie at the global free energy minimum. (c) Molecular reconstructions using the trained STAR pipeline of five representative points A-E selected from the testing $h2t(t)$ time series. Reconstructions $\hat{\mathbf{r}}$ (red) are superposed on the corresponding true configurations $\mathbf{r}$ (blue) extracted directly from the MD simulation. The head-to-tail distance of the true configuration and the RMSD under translational, rotational, and mirror alignment between the true and reconstructed configurations are reported under each image.  The STAR prediction of the location of each point on the smFES is illustrated in panel b and the corresponding dimensionless free energy $\beta F$ reported under each image.} \label{fig:Chig}
\end{figure*}

Application of the trained Chignolin STAR model is illustrated in Fig.~\ref{fig:Chig}. The heavy atom reconstruction accuracy over the 5 $\mu$s training trajectory is RMSD$_\mathrm{train}$ = 0.12 nm, that is only slightly diminished to RMSD$_\mathrm{test}$ =  0.14 nm over the 15 $\mu$s test trajectory, again illustrating good generalizability of the trained model. We illustrate for five selected points A-E in the $h2t(t)$ time trace (Fig.~\ref{fig:Chig}a) their projection onto the smFES $\beta F(\psi_1, \psi_2)$ (Fig.~\ref{fig:Chig}b) and their reconstructed $\hat{\mathbf{r}}(t) \in \mathbb{R}^{279}$ and true $\mathbf{r}(t) \in \mathbb{R}^{279}$ molecular structures (Fig.~\ref{fig:Chig}c). \blauw{Movie S2} presents an animation of molecular reconstructions and smFES projections for all data in the time series in Fig.~\ref{fig:Chig}a.

Configurations A and D correspond to the native hairpin state of the protein in which all native hydrogen bonds are intact and which lie within the deep global free energy minimum. The reconstruction accuracy of the densely sampled native fold is extremely good as indicated by the RMSD $\approx$ 0.05 nm. Configurations B, C, and E represent a sampling of the unfolded ensemble containing a diversity of random coiled states with some or all of the native hydrogen bonds broken. Despite the relatively sparser sampling and larger configurational diversity of the unfolded ensemble, the RMSD reconstruction accuracy is still better than 0.2 nm. This indicates that the training data, despite containing only four folding/unfolding transitions, provides a sufficiently dense and representative sampling of configurational space to enable accurate reconstruction of even transiently visited molecular conformations. It is unsurprising that configurations lying at high free energies (e.g., Configuration E, $\beta F$ = 10.5, RMSD = 0.15 nm) that are very sparsely sampled in the training data have poorer reconstruction accuracies than the densely sampled native configurations (e.g., Configuration A, $\beta F$ = 0.0, RMSD = 0.06 nm). We propose that adaptive sampling techniques to perform targeted sampling of infrequently-visited regions of the manifold may be beneficial in providing more training data to the STAR pipeline and improving the reconstruction accuracy of higher free energy configurations. Configurations B ($h2t$ = 2.1 nm) and C ($h2t$ = 2.4 nm) possess similar values of the $h2t$ distance but constitute two configurationally and thermodynamically distinct members of the Chignolin unfolded ensemble. Again, we observe that STAR accurately reconstructs configurations with similar instantaneous values of the scalar observable but which correspond to structurally different configurations that lie in very different regions of the smFES.

\section{Conclusions} \label{sec:concl}

This work presents the theoretical underpinnings and numerical implementation of an approach Single-molecule TAkens Reconstruction (STAR) to reconstruct molecular configurations from time series in a single experimentally-measurable observable. The basis for the approach rests upon the integration of Takens' Delay Embedding Theorem with tools from manifold learning, statistical thermodynamics, artificial neural networks, and rigid graph theory to extract a representation of the system state from the scalar time series and learn the \textit{a priori} unknown mapping to the molecular configuration from molecular dynamics simulation training data. The trained STAR model can then be applied to novel time series data to predict both the corresponding molecular configurations and their location and stability on the single molecule free energy surface. We have demonstrated and validated the approach in applications to a \ce{C_{24}H_{50}} polyethylene chain and the 10-residue engineered $\beta$-hairpin mini-protein Chignolin. In both cases we demonstrate that trained STAR models can robustly reconstruct the molecular configurations from time series data in the head-to-tail distance with RMSD accuracies better than 0.2 nm.

In this work, we adopt the head-to-tail distance as an experimental observable that can, in principle, be measured by a single molecule experimental technique such as smFRET. The head-to-tail time traces in this work are extracted from MD simulation trajectories in order to test our approach in applications where the ground truth molecular configurations are explicitly available from the simulations. These time traces can therefore be considered to represent synthetic and idealized smFRET data at arbitrarily high temporal resolution and subject to no measurement error or noise. The present work reports a computational proof-of-principle demonstration of STAR in this idealized limit. Applying STAR to real experimental time series must engage a number of concerns surrounding the experimental realities of smFRET measurements including millisecond limits in sampling frequency; shot noise, uncertainties, and unpredictable trajectory lengths due to photobleaching; degraded measurement reliabilities for fluorophores outside of the 2-8 nm range; and conformational perturbations induced by conjugation of the fluorescent probes \cite{Chang2011,Roy2008rr}. Furthermore, the STAR model must be trained and calibrated on MD training data and so the accuracy of the molecular potential functions and degree of sampling of the thermally-relevant configurational space will limit the quality to the trained model. A STAR model calibrated on MD simulation data employing a poor force field or which does not sample all of the experimentally-accessible states and transitions will not perform well when deployed on real experimental time series. In future work, we propose to engage these practical issues empirically by adding noise to the observables extracted from our simulation trajectories, limiting the accessible time resolution, limiting the length of the time series, and exploring the transferability of the trained models between molecular force fields. 

We would also like to explore technical innovations to determine optimal experimental observables (e.g., optimal FRET fluorophore placement) for high-accuracy molecular reconstruction \cite{Mittal2018}, the extension of STAR to multiplexed measurements \cite{Cao1998}, and the potential to reconstruct not just the molecular configuration but also the location and orientation of proximate solvent molecules using permutationally invariant representations of the solvent coordinates \cite{pietrucci2020novel,Han_2018}. It would also be of interest to explore the transferability of reconstructions learned under one set of conditions (e.g., temperature, pressure, salt concentration, mutations from wild type) to reconstruct those at another. We would also like to explore the possibility of adaptive sampling wherein the deployed model can identify regions of configurational space where it does not have sufficient training data to make accurate predictions and can conduct additional on-the-fly molecular simulations to supplement its training in these regions. This is anticipated to be particularly important for applications of STAR to large proteins where it is challenging to comprehensively sample the important configurational space. Finally, we also see potential applications of STAR in other applications where it is of interest to reconstruct the state of a dynamical system where it is challenging or impossible to obtain complete information on its state. As such, we envisage potential applications of the approach in fields such as climatology, epidemiology, and ecology.




\section*{Supplementary Material}

See supplementary material for simulation input files for the \ce{C_{24}H_{50}} molecular dynamics simulations, \blauw{Movie S1} showing an an animation of molecular reconstructions and smFES projections for \ce{C_{24}H_{50}}, \blauw{Movie S2} showing an an animation of molecular reconstructions and smFES projections for Chignolin.

\section*{Acknowledgments}

We thank Dr.~G{\"u}l H. Zerze for fruitful discussions. This material is based upon work supported by the National Science Foundation under Grant No.~DMS-1841810. We are grateful to D.E.~Shaw Research for sharing the Chignolin simulation trajectories.

\section*{Data Availability Statement}

Input files for the \ce{C_{24}H_{50}} molecular dynamics simulations are provided in the \blauw{Supplementary Material}. The Chignolin simulation trajectories reported in Ref.~\cite{deshaw} were obtained upon request from D.E.~Shaw Research.

\bibliography{bib}

\providecommand{\noopsort}[1]{}\providecommand{\singleletter}[1]{#1}%
\begin{thebibliography}{83}%
\makeatletter
\providecommand \@ifxundefined [1]{%
 \@ifx{#1\undefined}
}%
\providecommand \@ifnum [1]{%
 \ifnum #1\expandafter \@firstoftwo
 \else \expandafter \@secondoftwo
 \fi
}%
\providecommand \@ifx [1]{%
 \ifx #1\expandafter \@firstoftwo
 \else \expandafter \@secondoftwo
 \fi
}%
\providecommand \natexlab [1]{#1}%
\providecommand \enquote  [1]{``#1''}%
\providecommand \bibnamefont  [1]{#1}%
\providecommand \bibfnamefont [1]{#1}%
\providecommand \citenamefont [1]{#1}%
\providecommand \href@noop [0]{\@secondoftwo}%
\providecommand \href [0]{\begingroup \@sanitize@url \@href}%
\providecommand \@href[1]{\@@startlink{#1}\@@href}%
\providecommand \@@href[1]{\endgroup#1\@@endlink}%
\providecommand \@sanitize@url [0]{\catcode `\\12\catcode `\$12\catcode
  `\&12\catcode `\#12\catcode `\^12\catcode `\_12\catcode `\%12\relax}%
\providecommand \@@startlink[1]{}%
\providecommand \@@endlink[0]{}%
\providecommand \url  [0]{\begingroup\@sanitize@url \@url }%
\providecommand \@url [1]{\endgroup\@href {#1}{\urlprefix }}%
\providecommand \urlprefix  [0]{URL }%
\providecommand \Eprint [0]{\href }%
\providecommand \doibase [0]{http://dx.doi.org/}%
\providecommand \selectlanguage [0]{\@gobble}%
\providecommand \bibinfo  [0]{\@secondoftwo}%
\providecommand \bibfield  [0]{\@secondoftwo}%
\providecommand \translation [1]{[#1]}%
\providecommand \BibitemOpen [0]{}%
\providecommand \bibitemStop [0]{}%
\providecommand \bibitemNoStop [0]{.\EOS\space}%
\providecommand \EOS [0]{\spacefactor3000\relax}%
\providecommand \BibitemShut  [1]{\csname bibitem#1\endcsname}%
\let\auto@bib@innerbib\@empty
\bibitem [{\citenamefont {Frenkel}\ and\ \citenamefont
  {Smit}(2002)}]{frenkel2001understanding}%
  \BibitemOpen
  \bibfield  {author} {\bibinfo {author} {\bibfnamefont {D.}~\bibnamefont
  {Frenkel}}\ and\ \bibinfo {author} {\bibfnamefont {B.}~\bibnamefont {Smit}},\
  }\href@noop {} {\emph {\bibinfo {title} {Understanding Molecular Simulation:
  From algorithms to applications}}},\ Vol.~\bibinfo {volume} {1}\ (\bibinfo
  {publisher} {Academic press},\ \bibinfo {year} {2002})\BibitemShut {NoStop}%
\bibitem [{\citenamefont {Sch{\"{u}}ttelkopf}\ and\ \citenamefont {van
  Aalten}(2004)}]{PRODRG}%
  \BibitemOpen
  \bibfield  {author} {\bibinfo {author} {\bibfnamefont {A.~W.}\ \bibnamefont
  {Sch{\"{u}}ttelkopf}}\ and\ \bibinfo {author} {\bibfnamefont {D.~M.~F.}\
  \bibnamefont {van Aalten}},\ }\bibfield  {title} {\enquote {\bibinfo {title}
  {{PRODRG}: A tool for high-throughput crystallography of protein-ligand
  complexes},}\ }\href {\doibase 10.1107/S0907444904011679} {\bibfield
  {journal} {\bibinfo  {journal} {Acta Crystallographica Section D}\ }\textbf
  {\bibinfo {volume} {60}},\ \bibinfo {pages} {1355--1363} (\bibinfo {year}
  {2004})}\BibitemShut {NoStop}%
\bibitem [{\citenamefont {Chang}\ and\ \citenamefont
  {Rosenthal}(2011)}]{Chang2011}%
  \BibitemOpen
  \bibfield  {author} {\bibinfo {author} {\bibfnamefont {J.~C.}\ \bibnamefont
  {Chang}}\ and\ \bibinfo {author} {\bibfnamefont {S.~J.}\ \bibnamefont
  {Rosenthal}},\ }\enquote {\bibinfo {title} {Real-time quantum dot tracking of
  single proteins},}\ in\ \href {\doibase 10.1007/978-1-61779-052-2_4} {\emph
  {\bibinfo {booktitle} {Biomedical Nanotechnology: Methods and Protocols}}},\
  \bibinfo {editor} {edited by\ \bibinfo {editor} {\bibfnamefont {S.~J.}\
  \bibnamefont {Hurst}}}\ (\bibinfo  {publisher} {Humana Press},\ \bibinfo
  {address} {Totowa, NJ},\ \bibinfo {year} {2011})\ pp.\ \bibinfo {pages}
  {51--62}\BibitemShut {NoStop}%
\bibitem [{\citenamefont {Roy}, \citenamefont {Hohng},\ and\ \citenamefont
  {Ha}(2008)}]{Roy2008rr}%
  \BibitemOpen
  \bibfield  {author} {\bibinfo {author} {\bibfnamefont {R.}~\bibnamefont
  {Roy}}, \bibinfo {author} {\bibfnamefont {S.}~\bibnamefont {Hohng}}, \ and\
  \bibinfo {author} {\bibfnamefont {T.}~\bibnamefont {Ha}},\ }\bibfield
  {title} {\enquote {\bibinfo {title} {A practical guide to single-molecule
  {FRET}},}\ }\href {http://dx.doi.org/10.1038/nmeth.1208} {\bibfield
  {journal} {\bibinfo  {journal} {Nature Methods}\ }\textbf {\bibinfo {volume}
  {5}},\ \bibinfo {pages} {507--516} (\bibinfo {year} {2008})}\BibitemShut
  {NoStop}%
\bibitem [{\citenamefont {Flynn}\ \emph {et~al.}(2010)\citenamefont {Flynn},
  \citenamefont {Hanson}, \citenamefont {Alber},\ and\ \citenamefont
  {Yang}}]{flynn2010dynamic}%
  \BibitemOpen
  \bibfield  {author} {\bibinfo {author} {\bibfnamefont {E.~M.}\ \bibnamefont
  {Flynn}}, \bibinfo {author} {\bibfnamefont {J.~A.}\ \bibnamefont {Hanson}},
  \bibinfo {author} {\bibfnamefont {T.}~\bibnamefont {Alber}}, \ and\ \bibinfo
  {author} {\bibfnamefont {H.}~\bibnamefont {Yang}},\ }\bibfield  {title}
  {\enquote {\bibinfo {title} {Dynamic active-site protection by the {M}.
  tuberculosis protein tyrosine phosphatase {P}tpb lid domain},}\ }\href@noop
  {} {\bibfield  {journal} {\bibinfo  {journal} {Journal of the American
  Chemical Society}\ }\textbf {\bibinfo {volume} {132}},\ \bibinfo {pages}
  {4772--4780} (\bibinfo {year} {2010})}\BibitemShut {NoStop}%
\bibitem [{\citenamefont {Zerze}, \citenamefont {Best},\ and\ \citenamefont
  {Mittal}(2014)}]{Zerze2014mo}%
  \BibitemOpen
  \bibfield  {author} {\bibinfo {author} {\bibfnamefont {G.~H.}\ \bibnamefont
  {Zerze}}, \bibinfo {author} {\bibfnamefont {R.~B.}\ \bibnamefont {Best}}, \
  and\ \bibinfo {author} {\bibfnamefont {J.}~\bibnamefont {Mittal}},\
  }\bibfield  {title} {\enquote {\bibinfo {title} {Modest influence of {FRET}
  chromophores on the properties of unfolded proteins},}\ }\href {\doibase
  10.1016/j.bpj.2014.07.071} {\bibfield  {journal} {\bibinfo  {journal}
  {Biophysical Journal}\ }\textbf {\bibinfo {volume} {107}},\ \bibinfo {pages}
  {1654--1660} (\bibinfo {year} {2014})}\BibitemShut {NoStop}%
\bibitem [{\citenamefont {Takens}(1981)}]{Takens}%
  \BibitemOpen
  \bibfield  {author} {\bibinfo {author} {\bibfnamefont {F.}~\bibnamefont
  {Takens}},\ }\bibfield  {title} {\enquote {\bibinfo {title} {{Detecting
  strange attractors in turbulence}},}\ }\href
  {http://link.springer.com/content/pdf/10.1007/BFb0091924.pdf} {\bibfield
  {journal} {\bibinfo  {journal} {Dynamical Systems and Turbulence}\ }\textbf
  {\bibinfo {volume} {898}},\ \bibinfo {pages} {366--381} (\bibinfo {year}
  {1981})}\BibitemShut {NoStop}%
\bibitem [{\citenamefont {Sauer}, \citenamefont {Yorke},\ and\ \citenamefont
  {Casdagli}(1991)}]{sauer1991embedology}%
  \BibitemOpen
  \bibfield  {author} {\bibinfo {author} {\bibfnamefont {T.}~\bibnamefont
  {Sauer}}, \bibinfo {author} {\bibfnamefont {J.~A.}\ \bibnamefont {Yorke}}, \
  and\ \bibinfo {author} {\bibfnamefont {M.}~\bibnamefont {Casdagli}},\
  }\bibfield  {title} {\enquote {\bibinfo {title} {Embedology},}\ }\href@noop
  {} {\bibfield  {journal} {\bibinfo  {journal} {Journal of Statistical
  Physics}\ }\textbf {\bibinfo {volume} {65}},\ \bibinfo {pages} {579--616}
  (\bibinfo {year} {1991})}\BibitemShut {NoStop}%
\bibitem [{\citenamefont {Packard}\ \emph {et~al.}(1980)\citenamefont
  {Packard}, \citenamefont {Crutchfield}, \citenamefont {Farmer},\ and\
  \citenamefont {Shaw}}]{packard1980}%
  \BibitemOpen
  \bibfield  {author} {\bibinfo {author} {\bibfnamefont {N.}~\bibnamefont
  {Packard}}, \bibinfo {author} {\bibfnamefont {J.}~\bibnamefont
  {Crutchfield}}, \bibinfo {author} {\bibfnamefont {J.}~\bibnamefont {Farmer}},
  \ and\ \bibinfo {author} {\bibfnamefont {R.}~\bibnamefont {Shaw}},\
  }\bibfield  {title} {\enquote {\bibinfo {title} {Geometry from a time
  series},}\ }\href
  {http://journals.aps.org/prl/abstract/10.1103/PhysRevLett.45.712} {\bibfield
  {journal} {\bibinfo  {journal} {Physical Review Letters}\ }\textbf {\bibinfo
  {volume} {45}},\ \bibinfo {pages} {712--716} (\bibinfo {year}
  {1980})}\BibitemShut {NoStop}%
\bibitem [{\citenamefont {Broomhead}\ and\ \citenamefont
  {King}(1986)}]{Broomhead1986}%
  \BibitemOpen
  \bibfield  {author} {\bibinfo {author} {\bibfnamefont {D.~S.}\ \bibnamefont
  {Broomhead}}\ and\ \bibinfo {author} {\bibfnamefont {G.~P.}\ \bibnamefont
  {King}},\ }\bibfield  {title} {\enquote {\bibinfo {title} {{Extracting
  qualitative dynamics from experimental data}},}\ }\href {\doibase
  http://dx.doi.org/10.1016/0167-2789(86)90031-X} {\bibfield  {journal}
  {\bibinfo  {journal} {Physica D: Nonlinear Phenomena}\ }\textbf {\bibinfo
  {volume} {20}},\ \bibinfo {pages} {217--236} (\bibinfo {year}
  {1986})}\BibitemShut {NoStop}%
\bibitem [{\citenamefont {Cao}, \citenamefont {Mees},\ and\ \citenamefont
  {Judd}(1998)}]{Cao1998}%
  \BibitemOpen
  \bibfield  {author} {\bibinfo {author} {\bibfnamefont {L.}~\bibnamefont
  {Cao}}, \bibinfo {author} {\bibfnamefont {A.}~\bibnamefont {Mees}}, \ and\
  \bibinfo {author} {\bibfnamefont {K.}~\bibnamefont {Judd}},\ }\bibfield
  {title} {\enquote {\bibinfo {title} {{Dynamics from multivariate time
  series}},}\ }\href
  {http://www.sciencedirect.com/science/article/pii/S0167278998001511}
  {\bibfield  {journal} {\bibinfo  {journal} {Physica D: Nonlinear Phenomena}\
  }\textbf {\bibinfo {volume} {121}},\ \bibinfo {pages} {75--88} (\bibinfo
  {year} {1998})}\BibitemShut {NoStop}%
\bibitem [{\citenamefont {Stark}\ \emph {et~al.}(2003)\citenamefont {Stark},
  \citenamefont {Broomhead}, \citenamefont {Davies},\ and\ \citenamefont
  {Huke}}]{stark2003delay}%
  \BibitemOpen
  \bibfield  {author} {\bibinfo {author} {\bibfnamefont {J.}~\bibnamefont
  {Stark}}, \bibinfo {author} {\bibfnamefont {D.~S.}\ \bibnamefont
  {Broomhead}}, \bibinfo {author} {\bibfnamefont {M.}~\bibnamefont {Davies}}, \
  and\ \bibinfo {author} {\bibfnamefont {J.}~\bibnamefont {Huke}},\ }\bibfield
  {title} {\enquote {\bibinfo {title} {Delay embeddings for forced systems.
  {II}. {S}tochastic forcing},}\ }\href@noop {} {\bibfield  {journal} {\bibinfo
   {journal} {Journal of Nonlinear Science}\ }\textbf {\bibinfo {volume}
  {13}},\ \bibinfo {pages} {519--577} (\bibinfo {year} {2003})}\BibitemShut
  {NoStop}%
\bibitem [{\citenamefont {Vialar}(2009)}]{complex1}%
  \BibitemOpen
  \bibfield  {author} {\bibinfo {author} {\bibfnamefont {T.}~\bibnamefont
  {Vialar}},\ }\href
  {http://www.mathworks.com/support/books/book49040.html?category=4&language=1}
  {\emph {\bibinfo {title} {Complex and Chaotic Nonlinear Dynamics: Advances in
  Economics and Finance, Mathematics and Statistics}}}\ (\bibinfo  {publisher}
  {Springer},\ \bibinfo {year} {2009})\BibitemShut {NoStop}%
\bibitem [{\citenamefont {Kantz}\ and\ \citenamefont
  {Schreiber}(2005)}]{holger}%
  \BibitemOpen
  \bibfield  {author} {\bibinfo {author} {\bibfnamefont {H.}~\bibnamefont
  {Kantz}}\ and\ \bibinfo {author} {\bibfnamefont {T.}~\bibnamefont
  {Schreiber}},\ }\href@noop {} {\emph {\bibinfo {title} {{Nonlinear Time
  Series Analysis}}}},\ \bibinfo {edition} {2nd}\ ed.\ (\bibinfo  {publisher}
  {Cambridge},\ \bibinfo {year} {2005})\BibitemShut {NoStop}%
\bibitem [{\citenamefont {Ye}\ \emph {et~al.}(2015)\citenamefont {Ye},
  \citenamefont {Beamish}, \citenamefont {Glaser}, \citenamefont {Grant},
  \citenamefont {Hsieh}, \citenamefont {Richards}, \citenamefont {Schnute},\
  and\ \citenamefont {Sugihara}}]{PMID:25733874}%
  \BibitemOpen
  \bibfield  {author} {\bibinfo {author} {\bibfnamefont {H.}~\bibnamefont
  {Ye}}, \bibinfo {author} {\bibfnamefont {R.~J.}\ \bibnamefont {Beamish}},
  \bibinfo {author} {\bibfnamefont {S.~M.}\ \bibnamefont {Glaser}}, \bibinfo
  {author} {\bibfnamefont {S.~C.~H.}\ \bibnamefont {Grant}}, \bibinfo {author}
  {\bibfnamefont {C.-H.}\ \bibnamefont {Hsieh}}, \bibinfo {author}
  {\bibfnamefont {L.~J.}\ \bibnamefont {Richards}}, \bibinfo {author}
  {\bibfnamefont {J.~T.}\ \bibnamefont {Schnute}}, \ and\ \bibinfo {author}
  {\bibfnamefont {G.}~\bibnamefont {Sugihara}},\ }\bibfield  {title} {\enquote
  {\bibinfo {title} {Equation-free mechanistic ecosystem forecasting using
  empirical dynamic modeling},}\ }\href {\doibase 10.1073/pnas.1417063112}
  {\bibfield  {journal} {\bibinfo  {journal} {Proceedings of the National
  Academy of Sciences of the United States of America}\ }\textbf {\bibinfo
  {volume} {112}},\ \bibinfo {pages} {E1569---76} (\bibinfo {year}
  {2015})}\BibitemShut {NoStop}%
\bibitem [{\citenamefont {Wang}\ and\ \citenamefont {Ferguson}(2016)}]{Ferg16}%
  \BibitemOpen
  \bibfield  {author} {\bibinfo {author} {\bibfnamefont {J.}~\bibnamefont
  {Wang}}\ and\ \bibinfo {author} {\bibfnamefont {A.}~\bibnamefont
  {Ferguson}},\ }\bibfield  {title} {\enquote {\bibinfo {title} {Nonlinear
  reconstruction of single-molecule free energy surfaces from univariate time
  series},}\ }\href@noop {} {\bibfield  {journal} {\bibinfo  {journal}
  {Physical Review E}\ }\textbf {\bibinfo {volume} {93}},\ \bibinfo {pages}
  {032412} (\bibinfo {year} {2016})}\BibitemShut {NoStop}%
\bibitem [{\citenamefont {Wang}\ and\ \citenamefont
  {Ferguson}(2018{\natexlab{a}})}]{Ferg18}%
  \BibitemOpen
  \bibfield  {author} {\bibinfo {author} {\bibfnamefont {J.}~\bibnamefont
  {Wang}}\ and\ \bibinfo {author} {\bibfnamefont {A.}~\bibnamefont
  {Ferguson}},\ }\bibfield  {title} {\enquote {\bibinfo {title} {Recovery of
  protein folding funnels from single-molecule time series by delay embeddings
  and manifold learning},}\ }\href@noop {} {\bibfield  {journal} {\bibinfo
  {journal} {Journal of Physical Chemistry B}\ }\textbf {\bibinfo {volume} {122
  50}},\ \bibinfo {pages} {11931--11952} (\bibinfo {year}
  {2018}{\natexlab{a}})}\BibitemShut {NoStop}%
\bibitem [{\citenamefont {Humphrey}, \citenamefont {Dalke},\ and\ \citenamefont
  {Schulten}(1996)}]{humphrey1996vmd}%
  \BibitemOpen
  \bibfield  {author} {\bibinfo {author} {\bibfnamefont {W.}~\bibnamefont
  {Humphrey}}, \bibinfo {author} {\bibfnamefont {A.}~\bibnamefont {Dalke}}, \
  and\ \bibinfo {author} {\bibfnamefont {K.}~\bibnamefont {Schulten}},\
  }\bibfield  {title} {\enquote {\bibinfo {title} {{VMD}: {V}isual molecular
  dynamics},}\ }\href@noop {} {\bibfield  {journal} {\bibinfo  {journal}
  {Journal of Molecular Graphics}\ }\textbf {\bibinfo {volume} {14}},\ \bibinfo
  {pages} {33--38} (\bibinfo {year} {1996})}\BibitemShut {NoStop}%
\bibitem [{\citenamefont {Ferguson}\ \emph {et~al.}(2010)\citenamefont
  {Ferguson}, \citenamefont {Panagiotopoulos}, \citenamefont {Debenedetti},\
  and\ \citenamefont {Kevrekidis}}]{ferguson2010systematic}%
  \BibitemOpen
  \bibfield  {author} {\bibinfo {author} {\bibfnamefont {A.~L.}\ \bibnamefont
  {Ferguson}}, \bibinfo {author} {\bibfnamefont {A.~Z.}\ \bibnamefont
  {Panagiotopoulos}}, \bibinfo {author} {\bibfnamefont {P.~G.}\ \bibnamefont
  {Debenedetti}}, \ and\ \bibinfo {author} {\bibfnamefont {I.~G.}\ \bibnamefont
  {Kevrekidis}},\ }\bibfield  {title} {\enquote {\bibinfo {title} {Systematic
  determination of order parameters for chain dynamics using diffusion maps},}\
  }\href@noop {} {\bibfield  {journal} {\bibinfo  {journal} {Proceedings of the
  National Academy of Sciences of the United States of America}\ }\textbf
  {\bibinfo {volume} {107}},\ \bibinfo {pages} {13597--13602} (\bibinfo {year}
  {2010})}\BibitemShut {NoStop}%
\bibitem [{\citenamefont {Garc{\'\i}a}(1992)}]{garcia1992large}%
  \BibitemOpen
  \bibfield  {author} {\bibinfo {author} {\bibfnamefont {A.~E.}\ \bibnamefont
  {Garc{\'\i}a}},\ }\bibfield  {title} {\enquote {\bibinfo {title}
  {Large-amplitude nonlinear motions in proteins},}\ }\href@noop {} {\bibfield
  {journal} {\bibinfo  {journal} {Physical Review Letters}\ }\textbf {\bibinfo
  {volume} {68}},\ \bibinfo {pages} {2696} (\bibinfo {year}
  {1992})}\BibitemShut {NoStop}%
\bibitem [{\citenamefont {Amadei}, \citenamefont {Linssen},\ and\ \citenamefont
  {Berendsen}(1993)}]{amadei1993essential}%
  \BibitemOpen
  \bibfield  {author} {\bibinfo {author} {\bibfnamefont {A.}~\bibnamefont
  {Amadei}}, \bibinfo {author} {\bibfnamefont {A.}~\bibnamefont {Linssen}}, \
  and\ \bibinfo {author} {\bibfnamefont {H.~J.}\ \bibnamefont {Berendsen}},\
  }\bibfield  {title} {\enquote {\bibinfo {title} {Essential dynamics of
  proteins},}\ }\href@noop {} {\bibfield  {journal} {\bibinfo  {journal}
  {Proteins: Structure, Function, and Bioinformatics}\ }\textbf {\bibinfo
  {volume} {17}},\ \bibinfo {pages} {412--425} (\bibinfo {year}
  {1993})}\BibitemShut {NoStop}%
\bibitem [{\citenamefont {Hegger}\ \emph {et~al.}(2007)\citenamefont {Hegger},
  \citenamefont {Altis}, \citenamefont {Nguyen},\ and\ \citenamefont
  {Stock}}]{hegger2007complex}%
  \BibitemOpen
  \bibfield  {author} {\bibinfo {author} {\bibfnamefont {R.}~\bibnamefont
  {Hegger}}, \bibinfo {author} {\bibfnamefont {A.}~\bibnamefont {Altis}},
  \bibinfo {author} {\bibfnamefont {P.~H.}\ \bibnamefont {Nguyen}}, \ and\
  \bibinfo {author} {\bibfnamefont {G.}~\bibnamefont {Stock}},\ }\bibfield
  {title} {\enquote {\bibinfo {title} {How complex is the dynamics of peptide
  folding?}}\ }\href@noop {} {\bibfield  {journal} {\bibinfo  {journal}
  {Physical Review Letters}\ }\textbf {\bibinfo {volume} {98}},\ \bibinfo
  {pages} {028102} (\bibinfo {year} {2007})}\BibitemShut {NoStop}%
\bibitem [{\citenamefont {Zhuravlev}, \citenamefont {Materese},\ and\
  \citenamefont {Papoian}(2009)}]{Zhuravlev2009}%
  \BibitemOpen
  \bibfield  {author} {\bibinfo {author} {\bibfnamefont {P.~I.}\ \bibnamefont
  {Zhuravlev}}, \bibinfo {author} {\bibfnamefont {C.~K.}\ \bibnamefont
  {Materese}}, \ and\ \bibinfo {author} {\bibfnamefont {G.~A.}\ \bibnamefont
  {Papoian}},\ }\bibfield  {title} {\enquote {\bibinfo {title} {{Deconstructing
  the native state: Energy landscapes, function, and dynamics of globular
  proteins}},}\ }\href {\doibase 10.1021/jp810659u} {\bibfield  {journal}
  {\bibinfo  {journal} {Journal of Physical Chemistry B}\ }\textbf {\bibinfo
  {volume} {113}},\ \bibinfo {pages} {8800--8812} (\bibinfo {year}
  {2009})}\BibitemShut {NoStop}%
\bibitem [{\citenamefont {Das}\ \emph {et~al.}(2006)\citenamefont {Das},
  \citenamefont {Moll}, \citenamefont {Stamati}, \citenamefont {Kavraki},\ and\
  \citenamefont {Clementi}}]{das2006low}%
  \BibitemOpen
  \bibfield  {author} {\bibinfo {author} {\bibfnamefont {P.}~\bibnamefont
  {Das}}, \bibinfo {author} {\bibfnamefont {M.}~\bibnamefont {Moll}}, \bibinfo
  {author} {\bibfnamefont {H.}~\bibnamefont {Stamati}}, \bibinfo {author}
  {\bibfnamefont {L.~E.}\ \bibnamefont {Kavraki}}, \ and\ \bibinfo {author}
  {\bibfnamefont {C.}~\bibnamefont {Clementi}},\ }\bibfield  {title} {\enquote
  {\bibinfo {title} {Low-dimensional, free-energy landscapes of protein-folding
  reactions by nonlinear dimensionality reduction},}\ }\href@noop {} {\bibfield
   {journal} {\bibinfo  {journal} {Proceedings of the National Academy of
  Sciences of the United States of America}\ }\textbf {\bibinfo {volume}
  {103}},\ \bibinfo {pages} {9885--9890} (\bibinfo {year} {2006})}\BibitemShut
  {NoStop}%
\bibitem [{\citenamefont {Belkin}\ and\ \citenamefont
  {Niyogi}(2003)}]{belkin2003laplacian}%
  \BibitemOpen
  \bibfield  {author} {\bibinfo {author} {\bibfnamefont {M.}~\bibnamefont
  {Belkin}}\ and\ \bibinfo {author} {\bibfnamefont {P.}~\bibnamefont
  {Niyogi}},\ }\bibfield  {title} {\enquote {\bibinfo {title} {Laplacian
  eigenmaps for dimensionality reduction and data representation},}\
  }\href@noop {} {\bibfield  {journal} {\bibinfo  {journal} {Neural
  Computation}\ }\textbf {\bibinfo {volume} {15}},\ \bibinfo {pages}
  {1373--1396} (\bibinfo {year} {2003})}\BibitemShut {NoStop}%
\bibitem [{\citenamefont {Coifman}\ \emph
  {et~al.}(2008{\natexlab{a}})\citenamefont {Coifman}, \citenamefont
  {Kevrekidis}, \citenamefont {Lafon}, \citenamefont {Maggioni},\ and\
  \citenamefont {Nadler}}]{coifman2008diffusion}%
  \BibitemOpen
  \bibfield  {author} {\bibinfo {author} {\bibfnamefont {R.~R.}\ \bibnamefont
  {Coifman}}, \bibinfo {author} {\bibfnamefont {I.~G.}\ \bibnamefont
  {Kevrekidis}}, \bibinfo {author} {\bibfnamefont {S.}~\bibnamefont {Lafon}},
  \bibinfo {author} {\bibfnamefont {M.}~\bibnamefont {Maggioni}}, \ and\
  \bibinfo {author} {\bibfnamefont {B.}~\bibnamefont {Nadler}},\ }\bibfield
  {title} {\enquote {\bibinfo {title} {Diffusion maps, reduction coordinates,
  and low dimensional representation of stochastic systems},}\ }\href@noop {}
  {\bibfield  {journal} {\bibinfo  {journal} {Multiscale Modeling \&
  Simulation}\ }\textbf {\bibinfo {volume} {7}},\ \bibinfo {pages} {842--864}
  (\bibinfo {year} {2008}{\natexlab{a}})}\BibitemShut {NoStop}%
\bibitem [{\citenamefont {Ferguson}\ \emph {et~al.}(2011)\citenamefont
  {Ferguson}, \citenamefont {Panagiotopoulos}, \citenamefont {Kevrekidis},\
  and\ \citenamefont {Debenedetti}}]{ferguson2011cpl}%
  \BibitemOpen
  \bibfield  {author} {\bibinfo {author} {\bibfnamefont {A.~L.}\ \bibnamefont
  {Ferguson}}, \bibinfo {author} {\bibfnamefont {A.~Z.}\ \bibnamefont
  {Panagiotopoulos}}, \bibinfo {author} {\bibfnamefont {I.~G.}\ \bibnamefont
  {Kevrekidis}}, \ and\ \bibinfo {author} {\bibfnamefont {P.~G.}\ \bibnamefont
  {Debenedetti}},\ }\bibfield  {title} {\enquote {\bibinfo {title} {Nonlinear
  dimensionality reduction in molecular simulation: The diffusion map
  approach},}\ }\href@noop {} {\bibfield  {journal} {\bibinfo  {journal}
  {Chemical Physics Letters}\ }\textbf {\bibinfo {volume} {509}},\ \bibinfo
  {pages} {1--11} (\bibinfo {year} {2011})}\BibitemShut {NoStop}%
\bibitem [{\citenamefont {Coifman}\ \emph {et~al.}(2005)\citenamefont
  {Coifman}, \citenamefont {Lafon}, \citenamefont {Lee}, \citenamefont
  {Maggioni}, \citenamefont {Nadler}, \citenamefont {Warner},\ and\
  \citenamefont {Zucker}}]{coifman2005geometric}%
  \BibitemOpen
  \bibfield  {author} {\bibinfo {author} {\bibfnamefont {R.~R.}\ \bibnamefont
  {Coifman}}, \bibinfo {author} {\bibfnamefont {S.}~\bibnamefont {Lafon}},
  \bibinfo {author} {\bibfnamefont {A.~B.}\ \bibnamefont {Lee}}, \bibinfo
  {author} {\bibfnamefont {M.}~\bibnamefont {Maggioni}}, \bibinfo {author}
  {\bibfnamefont {B.}~\bibnamefont {Nadler}}, \bibinfo {author} {\bibfnamefont
  {F.}~\bibnamefont {Warner}}, \ and\ \bibinfo {author} {\bibfnamefont {S.~W.}\
  \bibnamefont {Zucker}},\ }\bibfield  {title} {\enquote {\bibinfo {title}
  {Geometric diffusions as a tool for harmonic analysis and structure
  definition of data: Diffusion maps},}\ }\href@noop {} {\bibfield  {journal}
  {\bibinfo  {journal} {Proceedings of the National Academy of Sciences of the
  United States of America}\ }\textbf {\bibinfo {volume} {102}},\ \bibinfo
  {pages} {7426--7431} (\bibinfo {year} {2005})}\BibitemShut {NoStop}%
\bibitem [{\citenamefont {Coifman}\ and\ \citenamefont
  {Lafon}(2006)}]{coifman2006diffusion}%
  \BibitemOpen
  \bibfield  {author} {\bibinfo {author} {\bibfnamefont {R.~R.}\ \bibnamefont
  {Coifman}}\ and\ \bibinfo {author} {\bibfnamefont {S.}~\bibnamefont
  {Lafon}},\ }\bibfield  {title} {\enquote {\bibinfo {title} {Diffusion
  maps},}\ }\href@noop {} {\bibfield  {journal} {\bibinfo  {journal} {Applied
  and Computational Harmonic Analysis}\ }\textbf {\bibinfo {volume} {21}},\
  \bibinfo {pages} {5--30} (\bibinfo {year} {2006})}\BibitemShut {NoStop}%
\bibitem [{\citenamefont {Nadler}\ \emph
  {et~al.}(2006{\natexlab{a}})\citenamefont {Nadler}, \citenamefont {Lafon},
  \citenamefont {Coifman},\ and\ \citenamefont {Kevrekidis}}]{lpbeltrami}%
  \BibitemOpen
  \bibfield  {author} {\bibinfo {author} {\bibfnamefont {B.}~\bibnamefont
  {Nadler}}, \bibinfo {author} {\bibfnamefont {S.}~\bibnamefont {Lafon}},
  \bibinfo {author} {\bibfnamefont {R.~R.}\ \bibnamefont {Coifman}}, \ and\
  \bibinfo {author} {\bibfnamefont {I.~G.}\ \bibnamefont {Kevrekidis}},\
  }\bibfield  {title} {\enquote {\bibinfo {title} {Diffusion maps, spectral
  clustering and reaction coordinates of dynamical systems},}\ }\href@noop {}
  {\bibfield  {journal} {\bibinfo  {journal} {Applied and Computational
  Harmonic Analysis}\ }\textbf {\bibinfo {volume} {21}},\ \bibinfo {pages}
  {113--127} (\bibinfo {year} {2006}{\natexlab{a}})}\BibitemShut {NoStop}%
\bibitem [{\citenamefont {Nadler}\ \emph
  {et~al.}(2006{\natexlab{b}})\citenamefont {Nadler}, \citenamefont {Lafon},
  \citenamefont {Coifman},\ and\ \citenamefont
  {Kevrekidis}}]{nadler2006advances}%
  \BibitemOpen
  \bibfield  {author} {\bibinfo {author} {\bibfnamefont {B.}~\bibnamefont
  {Nadler}}, \bibinfo {author} {\bibfnamefont {S.}~\bibnamefont {Lafon}},
  \bibinfo {author} {\bibfnamefont {R.~R.}\ \bibnamefont {Coifman}}, \ and\
  \bibinfo {author} {\bibfnamefont {I.~G.}\ \bibnamefont {Kevrekidis}},\
  }\bibfield  {title} {\enquote {\bibinfo {title} {Diffusion maps, spectral
  clustering and eigenfunctions of {F}okker-{P}lanck operators},}\ }in\
  \href@noop {} {\emph {\bibinfo {booktitle} {Advances in Neural Information
  Processing Systems 18: Proceedings of the 2005 Conference (Neural Information
  Processing)}}}\ (\bibinfo  {publisher} {The MIT Press},\ \bibinfo {year}
  {2006})\ pp.\ \bibinfo {pages} {955--962}\BibitemShut {NoStop}%
\bibitem [{\citenamefont {Ferguson}(2017)}]{Ferguson_2017}%
  \BibitemOpen
  \bibfield  {author} {\bibinfo {author} {\bibfnamefont {A.~L.}\ \bibnamefont
  {Ferguson}},\ }\bibfield  {title} {\enquote {\bibinfo {title} {Machine
  learning and data science in soft materials engineering},}\ }\href {\doibase
  10.1088/1361-648x/aa98bd} {\bibfield  {journal} {\bibinfo  {journal} {Journal
  of Physics: Condensed Matter}\ }\textbf {\bibinfo {volume} {30}},\ \bibinfo
  {pages} {043002} (\bibinfo {year} {2017})}\BibitemShut {NoStop}%
\bibitem [{\citenamefont {Sidky}, \citenamefont {Chen},\ and\ \citenamefont
  {Ferguson}(2020)}]{Sidky2020}%
  \BibitemOpen
  \bibfield  {author} {\bibinfo {author} {\bibfnamefont {H.}~\bibnamefont
  {Sidky}}, \bibinfo {author} {\bibfnamefont {W.}~\bibnamefont {Chen}}, \ and\
  \bibinfo {author} {\bibfnamefont {A.~L.}\ \bibnamefont {Ferguson}},\
  }\bibfield  {title} {\enquote {\bibinfo {title} {Machine learning for
  collective variable discovery and enhanced sampling in biomolecular
  simulation},}\ }\href {\doibase 10.1080/00268976.2020.1737742} {\bibfield
  {journal} {\bibinfo  {journal} {Molecular Physics}\ }\textbf {\bibinfo
  {volume} {118}},\ \bibinfo {pages} {e1737742} (\bibinfo {year}
  {2020})}\BibitemShut {NoStop}%
\bibitem [{\citenamefont {Coifman}\ \emph
  {et~al.}(2008{\natexlab{b}})\citenamefont {Coifman}, \citenamefont
  {Shkolnisky}, \citenamefont {Sigworth},\ and\ \citenamefont
  {Singer}}]{coifman2008graph}%
  \BibitemOpen
  \bibfield  {author} {\bibinfo {author} {\bibfnamefont {R.~R.}\ \bibnamefont
  {Coifman}}, \bibinfo {author} {\bibfnamefont {Y.}~\bibnamefont {Shkolnisky}},
  \bibinfo {author} {\bibfnamefont {F.~J.}\ \bibnamefont {Sigworth}}, \ and\
  \bibinfo {author} {\bibfnamefont {A.}~\bibnamefont {Singer}},\ }\bibfield
  {title} {\enquote {\bibinfo {title} {Graph {L}aplacian tomography from
  unknown random projections},}\ }\href@noop {} {\bibfield  {journal} {\bibinfo
   {journal} {IEEE Transactions on Image Processing}\ }\textbf {\bibinfo
  {volume} {17}},\ \bibinfo {pages} {1891--1899} (\bibinfo {year}
  {2008}{\natexlab{b}})}\BibitemShut {NoStop}%
\bibitem [{\citenamefont {Wang}\ and\ \citenamefont
  {Ferguson}(2018{\natexlab{b}})}]{Wang2017}%
  \BibitemOpen
  \bibfield  {author} {\bibinfo {author} {\bibfnamefont {J.}~\bibnamefont
  {Wang}}\ and\ \bibinfo {author} {\bibfnamefont {A.~L.}\ \bibnamefont
  {Ferguson}},\ }\bibfield  {title} {\enquote {\bibinfo {title} {Nonlinear
  machine learning in simulations of soft and biological materials},}\ }\href
  {\doibase 10.1080/08927022.2017.1400164} {\bibfield  {journal} {\bibinfo
  {journal} {Molecular Simulation}\ }\textbf {\bibinfo {volume} {44}},\
  \bibinfo {pages} {1090--1107} (\bibinfo {year}
  {2018}{\natexlab{b}})}\BibitemShut {NoStop}%
\bibitem [{\citenamefont {Tenenbaum}, \citenamefont {de~Silva},\ and\
  \citenamefont {Langford}(2000)}]{Tenenbaum2000}%
  \BibitemOpen
  \bibfield  {author} {\bibinfo {author} {\bibfnamefont {J.~B.}\ \bibnamefont
  {Tenenbaum}}, \bibinfo {author} {\bibfnamefont {V.}~\bibnamefont {de~Silva}},
  \ and\ \bibinfo {author} {\bibfnamefont {J.~C.}\ \bibnamefont {Langford}},\
  }\bibfield  {title} {\enquote {\bibinfo {title} {{A Global Geometric
  Framework for Nonlinear Dimensionality Reduction}},}\ }\href {\doibase
  10.1126/science.290.5500.2319} {\bibfield  {journal} {\bibinfo  {journal}
  {Science}\ }\textbf {\bibinfo {volume} {290}},\ \bibinfo {pages} {2319--2323}
  (\bibinfo {year} {2000})}\BibitemShut {NoStop}%
\bibitem [{\citenamefont {Weinberger}\ and\ \citenamefont
  {Saul}(2006)}]{Saul2006}%
  \BibitemOpen
  \bibfield  {author} {\bibinfo {author} {\bibfnamefont {K.~Q.}\ \bibnamefont
  {Weinberger}}\ and\ \bibinfo {author} {\bibfnamefont {L.~K.}\ \bibnamefont
  {Saul}},\ }\bibfield  {title} {\enquote {\bibinfo {title} {Unsupervised
  learning of image manifolds by semidefinite programming},}\ }\href@noop {}
  {\bibfield  {journal} {\bibinfo  {journal} {International Journal of Computer
  Vision volume}\ }\textbf {\bibinfo {volume} {70}},\ \bibinfo {pages} {77--90}
  (\bibinfo {year} {2006})}\BibitemShut {NoStop}%
\bibitem [{\citenamefont {Li}\ \emph {et~al.}(2006)\citenamefont {Li},
  \citenamefont {Guo}, \citenamefont {Chen}, \citenamefont {Nie},\ and\
  \citenamefont {Yang}}]{Li2006}%
  \BibitemOpen
  \bibfield  {author} {\bibinfo {author} {\bibfnamefont {C.-G.}\ \bibnamefont
  {Li}}, \bibinfo {author} {\bibfnamefont {J.}~\bibnamefont {Guo}}, \bibinfo
  {author} {\bibfnamefont {G.}~\bibnamefont {Chen}}, \bibinfo {author}
  {\bibfnamefont {X.-F.}\ \bibnamefont {Nie}}, \ and\ \bibinfo {author}
  {\bibfnamefont {Z.}~\bibnamefont {Yang}},\ }\bibfield  {title} {\enquote
  {\bibinfo {title} {A version of {I}somap with explicit mapping},}\ }in\
  \href@noop {} {\emph {\bibinfo {booktitle} {IEEE International Conference on
  Machine Learning and Cybernetics}}}\ (\bibinfo {year} {2006})\ pp.\ \bibinfo
  {pages} {3201---3206}\BibitemShut {NoStop}%
\bibitem [{\citenamefont {Wang}(2011)}]{Jwang2011}%
  \BibitemOpen
  \bibfield  {author} {\bibinfo {author} {\bibfnamefont {J.}~\bibnamefont
  {Wang}},\ }\href@noop {} {\emph {\bibinfo {title} {Geometric Structure of
  High-Dimensional Data and Dimensionality Reduction}}}\ (\bibinfo  {publisher}
  {Springer},\ \bibinfo {year} {2011})\BibitemShut {NoStop}%
\bibitem [{\citenamefont {Roweis}\ and\ \citenamefont
  {Saul}(2000)}]{roweis2000nonlinear}%
  \BibitemOpen
  \bibfield  {author} {\bibinfo {author} {\bibfnamefont {S.~T.}\ \bibnamefont
  {Roweis}}\ and\ \bibinfo {author} {\bibfnamefont {L.~K.}\ \bibnamefont
  {Saul}},\ }\bibfield  {title} {\enquote {\bibinfo {title} {Nonlinear
  dimensionality reduction by locally linear embedding},}\ }\href@noop {}
  {\bibfield  {journal} {\bibinfo  {journal} {Science}\ }\textbf {\bibinfo
  {volume} {290}},\ \bibinfo {pages} {2323--2326} (\bibinfo {year}
  {2000})}\BibitemShut {NoStop}%
\bibitem [{\citenamefont {Zhang}\ and\ \citenamefont {Wang}(2006)}]{Zhang2006}%
  \BibitemOpen
  \bibfield  {author} {\bibinfo {author} {\bibfnamefont {Z.}~\bibnamefont
  {Zhang}}\ and\ \bibinfo {author} {\bibfnamefont {J.}~\bibnamefont {Wang}},\
  }\bibfield  {title} {\enquote {\bibinfo {title} {Mlle: Modified locally
  linear embedding using multiple weights},}\ }in\ \href@noop {} {\emph
  {\bibinfo {booktitle} {Proceedings of the 19th International Conference on
  Neural Information Processing Systems}}},\ \bibinfo {series and number}
  {NIPS'06}\ (\bibinfo  {publisher} {MIT Press},\ \bibinfo {address}
  {Cambridge, MA, USA},\ \bibinfo {year} {2006})\ pp.\ \bibinfo {pages}
  {1593--1600}\BibitemShut {NoStop}%
\bibitem [{\citenamefont {Kabsch}(1976)}]{kabsch1976solution}%
  \BibitemOpen
  \bibfield  {author} {\bibinfo {author} {\bibfnamefont {W.}~\bibnamefont
  {Kabsch}},\ }\bibfield  {title} {\enquote {\bibinfo {title} {A solution for
  the best rotation to relate two sets of vectors},}\ }\href@noop {} {\bibfield
   {journal} {\bibinfo  {journal} {Acta Crystallographica Section A: Crystal
  Physics, Diffraction, Theoretical and General Crystallography}\ }\textbf
  {\bibinfo {volume} {32}},\ \bibinfo {pages} {922--923} (\bibinfo {year}
  {1976})}\BibitemShut {NoStop}%
\bibitem [{\citenamefont {Sonday}, \citenamefont {Haataja},\ and\ \citenamefont
  {Kevrekidis}(2009)}]{sonday2009coarse}%
  \BibitemOpen
  \bibfield  {author} {\bibinfo {author} {\bibfnamefont {B.~E.}\ \bibnamefont
  {Sonday}}, \bibinfo {author} {\bibfnamefont {M.}~\bibnamefont {Haataja}}, \
  and\ \bibinfo {author} {\bibfnamefont {I.~G.}\ \bibnamefont {Kevrekidis}},\
  }\bibfield  {title} {\enquote {\bibinfo {title} {Coarse-graining the dynamics
  of a driven interface in the presence of mobile impurities: Effective
  description via diffusion maps},}\ }\href@noop {} {\bibfield  {journal}
  {\bibinfo  {journal} {Physical Review E}\ }\textbf {\bibinfo {volume} {80}},\
  \bibinfo {pages} {031102} (\bibinfo {year} {2009})}\BibitemShut {NoStop}%
\bibitem [{\citenamefont {Laing}, \citenamefont {Frewen},\ and\ \citenamefont
  {Kevrekidis}(2007)}]{laing2007coarse}%
  \BibitemOpen
  \bibfield  {author} {\bibinfo {author} {\bibfnamefont {C.}~\bibnamefont
  {Laing}}, \bibinfo {author} {\bibfnamefont {T.}~\bibnamefont {Frewen}}, \
  and\ \bibinfo {author} {\bibfnamefont {I.}~\bibnamefont {Kevrekidis}},\
  }\bibfield  {title} {\enquote {\bibinfo {title} {Coarse-grained dynamics of
  an activity bump in a neural field model},}\ }\href@noop {} {\bibfield
  {journal} {\bibinfo  {journal} {Nonlinearity}\ }\textbf {\bibinfo {volume}
  {20}},\ \bibinfo {pages} {2127} (\bibinfo {year} {2007})}\BibitemShut
  {NoStop}%
\bibitem [{\citenamefont {Long}\ and\ \citenamefont
  {Ferguson}(2019)}]{long2019landmark}%
  \BibitemOpen
  \bibfield  {author} {\bibinfo {author} {\bibfnamefont {A.~W.}\ \bibnamefont
  {Long}}\ and\ \bibinfo {author} {\bibfnamefont {A.~L.}\ \bibnamefont
  {Ferguson}},\ }\bibfield  {title} {\enquote {\bibinfo {title} {Landmark
  diffusion maps ({L}-dmaps): {A}ccelerated manifold learning out-of-sample
  extension},}\ }\href@noop {} {\bibfield  {journal} {\bibinfo  {journal}
  {Applied and Computational Harmonic Analysis}\ }\textbf {\bibinfo {volume}
  {47}},\ \bibinfo {pages} {190--211} (\bibinfo {year} {2019})}\BibitemShut
  {NoStop}%
\bibitem [{\citenamefont {Letellier}\ \emph {et~al.}(1998)\citenamefont
  {Letellier}, \citenamefont {Maquet}, \citenamefont {Le~Sceller},
  \citenamefont {Gouesbet},\ and\ \citenamefont {Aguirre}}]{letellier1998non}%
  \BibitemOpen
  \bibfield  {author} {\bibinfo {author} {\bibfnamefont {C.}~\bibnamefont
  {Letellier}}, \bibinfo {author} {\bibfnamefont {J.}~\bibnamefont {Maquet}},
  \bibinfo {author} {\bibfnamefont {L.}~\bibnamefont {Le~Sceller}}, \bibinfo
  {author} {\bibfnamefont {G.}~\bibnamefont {Gouesbet}}, \ and\ \bibinfo
  {author} {\bibfnamefont {L.}~\bibnamefont {Aguirre}},\ }\bibfield  {title}
  {\enquote {\bibinfo {title} {On the non-equivalence of observables in
  phase-space reconstructions from recorded time series},}\ }\href@noop {}
  {\bibfield  {journal} {\bibinfo  {journal} {Journal of Physics A:
  Mathematical and General}\ }\textbf {\bibinfo {volume} {31}},\ \bibinfo
  {pages} {7913} (\bibinfo {year} {1998})}\BibitemShut {NoStop}%
\bibitem [{\citenamefont {Cross}\ and\ \citenamefont
  {Gilmore}(2010)}]{cross2010differential}%
  \BibitemOpen
  \bibfield  {author} {\bibinfo {author} {\bibfnamefont {D.~J.}\ \bibnamefont
  {Cross}}\ and\ \bibinfo {author} {\bibfnamefont {R.}~\bibnamefont
  {Gilmore}},\ }\bibfield  {title} {\enquote {\bibinfo {title} {Differential
  embedding of the {Lorenz} attractor},}\ }\href@noop {} {\bibfield  {journal}
  {\bibinfo  {journal} {Physical Review E}\ }\textbf {\bibinfo {volume} {81}},\
  \bibinfo {pages} {066220} (\bibinfo {year} {2010})}\BibitemShut {NoStop}%
\bibitem [{\citenamefont {Letellier}\ and\ \citenamefont
  {Gouesbet}(1996)}]{letellier1996topological}%
  \BibitemOpen
  \bibfield  {author} {\bibinfo {author} {\bibfnamefont {C.}~\bibnamefont
  {Letellier}}\ and\ \bibinfo {author} {\bibfnamefont {G.}~\bibnamefont
  {Gouesbet}},\ }\bibfield  {title} {\enquote {\bibinfo {title} {Topological
  characterization of reconstructed attractors modding out symmetries},}\
  }\href@noop {} {\bibfield  {journal} {\bibinfo  {journal} {Journal de
  Physique II}\ }\textbf {\bibinfo {volume} {6}},\ \bibinfo {pages}
  {1615--1638} (\bibinfo {year} {1996})}\BibitemShut {NoStop}%
\bibitem [{\citenamefont {Fraser}\ and\ \citenamefont
  {Swinney}(1986{\natexlab{a}})}]{Fraser86}%
  \BibitemOpen
  \bibfield  {author} {\bibinfo {author} {\bibfnamefont {A.~M.}\ \bibnamefont
  {Fraser}}\ and\ \bibinfo {author} {\bibfnamefont {H.~L.}\ \bibnamefont
  {Swinney}},\ }\bibfield  {title} {\enquote {\bibinfo {title} {Independent
  coordinates for strange attractors from mutual information},}\ }\href@noop {}
  {\bibfield  {journal} {\bibinfo  {journal} {Physical Review A}\ }\textbf
  {\bibinfo {volume} {33}},\ \bibinfo {pages} {1134} (\bibinfo {year}
  {1986}{\natexlab{a}})}\BibitemShut {NoStop}%
\bibitem [{\citenamefont {Fraser}\ and\ \citenamefont
  {Swinney}(1986{\natexlab{b}})}]{MI}%
  \BibitemOpen
  \bibfield  {author} {\bibinfo {author} {\bibfnamefont {A.~M.}\ \bibnamefont
  {Fraser}}\ and\ \bibinfo {author} {\bibfnamefont {H.~L.}\ \bibnamefont
  {Swinney}},\ }\bibfield  {title} {\enquote {\bibinfo {title} {Independent
  coordinates for strange attractors from mutual information},}\ }\href
  {\doibase 10.1103/PhysRevA.33.1134} {\bibfield  {journal} {\bibinfo
  {journal} {Physical Review A}\ }\textbf {\bibinfo {volume} {33}},\ \bibinfo
  {pages} {1134--1140} (\bibinfo {year} {1986}{\natexlab{b}})}\BibitemShut
  {NoStop}%
\bibitem [{\citenamefont {Cao}(1997)}]{Cao97}%
  \BibitemOpen
  \bibfield  {author} {\bibinfo {author} {\bibfnamefont {L.}~\bibnamefont
  {Cao}},\ }\bibfield  {title} {\enquote {\bibinfo {title} {Practical method
  for determining the minimum embedding dimension of a scalar time series},}\
  }\href {\doibase http://dx.doi.org/10.1016/S0167-2789(97)00118-8} {\bibfield
  {journal} {\bibinfo  {journal} {Physica D: Nonlinear Phenomena}\ }\textbf
  {\bibinfo {volume} {110}},\ \bibinfo {pages} {43--50} (\bibinfo {year}
  {1997})}\BibitemShut {NoStop}%
\bibitem [{\citenamefont {Villani}\ and\ \citenamefont {{Zaldivar
  Comenges}}(2000)}]{VillaniB}%
  \BibitemOpen
  \bibfield  {author} {\bibinfo {author} {\bibfnamefont {V.}~\bibnamefont
  {Villani}}\ and\ \bibinfo {author} {\bibfnamefont {J.~M.}\ \bibnamefont
  {{Zaldivar Comenges}}},\ }\bibfield  {title} {\enquote {\bibinfo {title}
  {{Analysis of biomolecular chaos in aqueous solution}},}\ }\href {\doibase
  10.1007/s002140000121} {\bibfield  {journal} {\bibinfo  {journal}
  {Theoretical Chemistry Accounts: Theory, Computation, and Modeling
  (Theoretica Chimica Acta)}\ }\textbf {\bibinfo {volume} {104}},\ \bibinfo
  {pages} {290--295} (\bibinfo {year} {2000})}\BibitemShut {NoStop}%
\bibitem [{\citenamefont {Kennel}, \citenamefont {Brown},\ and\ \citenamefont
  {Abarbanel}(1992)}]{kennel1992determining}%
  \BibitemOpen
  \bibfield  {author} {\bibinfo {author} {\bibfnamefont {M.~B.}\ \bibnamefont
  {Kennel}}, \bibinfo {author} {\bibfnamefont {R.}~\bibnamefont {Brown}}, \
  and\ \bibinfo {author} {\bibfnamefont {H.~D.}\ \bibnamefont {Abarbanel}},\
  }\bibfield  {title} {\enquote {\bibinfo {title} {Determining embedding
  dimension for phase-space reconstruction using a geometrical construction},}\
  }\href@noop {} {\bibfield  {journal} {\bibinfo  {journal} {Physical Review
  A}\ }\textbf {\bibinfo {volume} {45}},\ \bibinfo {pages} {3403} (\bibinfo
  {year} {1992})}\BibitemShut {NoStop}%
\bibitem [{\citenamefont {Stark}(1999)}]{stark1999delay}%
  \BibitemOpen
  \bibfield  {author} {\bibinfo {author} {\bibfnamefont {J.}~\bibnamefont
  {Stark}},\ }\bibfield  {title} {\enquote {\bibinfo {title} {Delay embeddings
  for forced systems. {I}. {D}eterministic forcing},}\ }\href@noop {}
  {\bibfield  {journal} {\bibinfo  {journal} {Journal of Nonlinear Science}\
  }\textbf {\bibinfo {volume} {9}},\ \bibinfo {pages} {255--332} (\bibinfo
  {year} {1999})}\BibitemShut {NoStop}%
\bibitem [{\citenamefont {Cover}\ and\ \citenamefont {Hart}(1967)}]{Cover1967}%
  \BibitemOpen
  \bibfield  {author} {\bibinfo {author} {\bibfnamefont {T.~M.}\ \bibnamefont
  {Cover}}\ and\ \bibinfo {author} {\bibfnamefont {P.~E.}\ \bibnamefont
  {Hart}},\ }\bibfield  {title} {\enquote {\bibinfo {title} {Nearest neighbor
  pattern classification},}\ }\href@noop {} {\bibfield  {journal} {\bibinfo
  {journal} {IEEE Transactions on Information Theory}\ }\textbf {\bibinfo
  {volume} {13}},\ \bibinfo {pages} {21--27} (\bibinfo {year}
  {1967})}\BibitemShut {NoStop}%
\bibitem [{\citenamefont {Sch{\"o}lkopf}\ and\ \citenamefont
  {Smola}(2002)}]{Scholkopf2002}%
  \BibitemOpen
  \bibfield  {author} {\bibinfo {author} {\bibfnamefont {B.}~\bibnamefont
  {Sch{\"o}lkopf}}\ and\ \bibinfo {author} {\bibfnamefont {A.}~\bibnamefont
  {Smola}},\ }\href@noop {} {\emph {\bibinfo {title} {Learning with Kernels:
  Support Vector Machines, Regularization, Optimization, and Beyond}}},\
  Adaptive Computation and Machine Learning\ (\bibinfo  {publisher} {MIT
  Press},\ \bibinfo {address} {Cambridge, MA, USA},\ \bibinfo {year}
  {2002})\BibitemShut {NoStop}%
\bibitem [{\citenamefont {Hauser}\ and\ \citenamefont
  {Ray}(2017)}]{Principlesriemannian}%
  \BibitemOpen
  \bibfield  {author} {\bibinfo {author} {\bibfnamefont {M.}~\bibnamefont
  {Hauser}}\ and\ \bibinfo {author} {\bibfnamefont {A.}~\bibnamefont {Ray}},\
  }\bibfield  {title} {\enquote {\bibinfo {title} {Principles of {R}iemannian
  geometry in neural networks},}\ }in\ \href@noop {} {\emph {\bibinfo
  {booktitle} {Advances in Neural Information Processing Systems 30 (NIPS
  2017)}}},\ \bibinfo {editor} {edited by\ \bibinfo {editor} {\bibfnamefont
  {I.}~\bibnamefont {Guyon}}, \bibinfo {editor} {\bibfnamefont
  {U.}~\bibnamefont {Luxburg}}, \bibinfo {editor} {\bibfnamefont
  {S.}~\bibnamefont {Bengio}}, \bibinfo {editor} {\bibfnamefont
  {H.}~\bibnamefont {Wallach}}, \bibinfo {editor} {\bibfnamefont
  {R.}~\bibnamefont {Fergus}}, \bibinfo {editor} {\bibfnamefont
  {S.}~\bibnamefont {Vishwanathan}}, \ and\ \bibinfo {editor} {\bibfnamefont
  {R.}~\bibnamefont {Garnett}}}\ (\bibinfo {year} {2017})\ pp.\ \bibinfo
  {pages} {2808--2817}\BibitemShut {NoStop}%
\bibitem [{\citenamefont {Hassoun}\ \emph {et~al.}(1996)\citenamefont
  {Hassoun}, \citenamefont {Intrator}, \citenamefont {McKay},\ and\
  \citenamefont {Christian}}]{Hassoun1996}%
  \BibitemOpen
  \bibfield  {author} {\bibinfo {author} {\bibfnamefont {M.~H.}\ \bibnamefont
  {Hassoun}}, \bibinfo {author} {\bibfnamefont {N.}~\bibnamefont {Intrator}},
  \bibinfo {author} {\bibfnamefont {S.}~\bibnamefont {McKay}}, \ and\ \bibinfo
  {author} {\bibfnamefont {W.}~\bibnamefont {Christian}},\ }\bibfield  {title}
  {\enquote {\bibinfo {title} {Fundamentals of artificial neural networks},}\
  }\href {\doibase 10.1063/1.4822376} {\bibfield  {journal} {\bibinfo
  {journal} {Computers in Physics}\ }\textbf {\bibinfo {volume} {10}},\
  \bibinfo {pages} {137--137} (\bibinfo {year} {1996})}\BibitemShut {NoStop}%
\bibitem [{\citenamefont {Dokmanic}\ \emph
  {et~al.}(2015{\natexlab{a}})\citenamefont {Dokmanic}, \citenamefont
  {Parhizkar}, \citenamefont {Ranieri},\ and\ \citenamefont
  {Vetterli}}]{Euclid}%
  \BibitemOpen
  \bibfield  {author} {\bibinfo {author} {\bibfnamefont {I.}~\bibnamefont
  {Dokmanic}}, \bibinfo {author} {\bibfnamefont {R.}~\bibnamefont {Parhizkar}},
  \bibinfo {author} {\bibfnamefont {J.}~\bibnamefont {Ranieri}}, \ and\
  \bibinfo {author} {\bibfnamefont {M.}~\bibnamefont {Vetterli}},\ }\bibfield
  {title} {\enquote {\bibinfo {title} {Euclidean distance matrices: Essential
  theory, algorithms, and applications},}\ }\href {\doibase
  10.1109/MSP.2015.2398954} {\bibfield  {journal} {\bibinfo  {journal} {IEEE
  Signal Processing Magazine}\ }\textbf {\bibinfo {volume} {32}},\ \bibinfo
  {pages} {12--30} (\bibinfo {year} {2015}{\natexlab{a}})}\BibitemShut
  {NoStop}%
\bibitem [{\citenamefont {Singer}(2008)}]{Singer2008}%
  \BibitemOpen
  \bibfield  {author} {\bibinfo {author} {\bibfnamefont {A.}~\bibnamefont
  {Singer}},\ }\bibfield  {title} {\enquote {\bibinfo {title} {A remark on
  global positioning from local distances},}\ }\href {\doibase
  10.1073/pnas.0709842104} {\bibfield  {journal} {\bibinfo  {journal}
  {Proceedings of the National Academy of Sciences}\ }\textbf {\bibinfo
  {volume} {105}},\ \bibinfo {pages} {9507--9511} (\bibinfo {year}
  {2008})}\BibitemShut {NoStop}%
\bibitem [{\citenamefont {Dokmanic}\ \emph
  {et~al.}(2015{\natexlab{b}})\citenamefont {Dokmanic}, \citenamefont
  {Parhizkar}, \citenamefont {Ranieri},\ and\ \citenamefont
  {Vetterli}}]{Dokmanic_2015}%
  \BibitemOpen
  \bibfield  {author} {\bibinfo {author} {\bibfnamefont {I.}~\bibnamefont
  {Dokmanic}}, \bibinfo {author} {\bibfnamefont {R.}~\bibnamefont {Parhizkar}},
  \bibinfo {author} {\bibfnamefont {J.}~\bibnamefont {Ranieri}}, \ and\
  \bibinfo {author} {\bibfnamefont {M.}~\bibnamefont {Vetterli}},\ }\bibfield
  {title} {\enquote {\bibinfo {title} {Euclidean distance matrices: Essential
  theory, algorithms, and applications},}\ }\href {\doibase
  10.1109/msp.2015.2398954} {\bibfield  {journal} {\bibinfo  {journal} {IEEE
  Signal Processing Magazine}\ }\textbf {\bibinfo {volume} {32}},\ \bibinfo
  {pages} {12--30} (\bibinfo {year} {2015}{\natexlab{b}})}\BibitemShut
  {NoStop}%
\bibitem [{\citenamefont {Crippen}(1978)}]{CRIPPEN1978449}%
  \BibitemOpen
  \bibfield  {author} {\bibinfo {author} {\bibfnamefont {G.}~\bibnamefont
  {Crippen}},\ }\bibfield  {title} {\enquote {\bibinfo {title} {Note rapid
  calculation of coordinates from distance matrices},}\ }\href {\doibase
  https://doi.org/10.1016/0021-9991(78)90081-5} {\bibfield  {journal} {\bibinfo
   {journal} {Journal of Computational Physics}\ }\textbf {\bibinfo {volume}
  {26}},\ \bibinfo {pages} {449 -- 452} (\bibinfo {year} {1978})}\BibitemShut
  {NoStop}%
\bibitem [{\citenamefont {Sch{\"o}nemann}(1966)}]{Schnemann1966AGS}%
  \BibitemOpen
  \bibfield  {author} {\bibinfo {author} {\bibfnamefont {P.~H.}\ \bibnamefont
  {Sch{\"o}nemann}},\ }\bibfield  {title} {\enquote {\bibinfo {title} {A
  generalized solution of the orthogonal {P}rocrustes problem},}\ }\href@noop
  {} {\bibfield  {journal} {\bibinfo  {journal} {Psychometrika}\ }\textbf
  {\bibinfo {volume} {31}},\ \bibinfo {pages} {1--10} (\bibinfo {year}
  {1966})}\BibitemShut {NoStop}%
\bibitem [{\citenamefont {van~der Spoel}\ \emph {et~al.}(2005)\citenamefont
  {van~der Spoel}, \citenamefont {Lindahl}, \citenamefont {Hess}, \citenamefont
  {Groenhof}, \citenamefont {Mark},\ and\ \citenamefont {Berendsen}}]{Gromacs}%
  \BibitemOpen
  \bibfield  {author} {\bibinfo {author} {\bibfnamefont {D.}~\bibnamefont
  {van~der Spoel}}, \bibinfo {author} {\bibfnamefont {E.}~\bibnamefont
  {Lindahl}}, \bibinfo {author} {\bibfnamefont {B.}~\bibnamefont {Hess}},
  \bibinfo {author} {\bibfnamefont {G.}~\bibnamefont {Groenhof}}, \bibinfo
  {author} {\bibfnamefont {A.~E.}\ \bibnamefont {Mark}}, \ and\ \bibinfo
  {author} {\bibfnamefont {H.~J.~C.}\ \bibnamefont {Berendsen}},\ }\bibfield
  {title} {\enquote {\bibinfo {title} {{GROMACS}: Fast, flexible, and free},}\
  }\href {\doibase 10.1002/jcc.20291} {\bibfield  {journal} {\bibinfo
  {journal} {Journal of Computational Chemistry}\ }\textbf {\bibinfo {volume}
  {26}},\ \bibinfo {pages} {1701--1718} (\bibinfo {year} {2005})}\BibitemShut
  {NoStop}%
\bibitem [{\citenamefont {Martin}\ and\ \citenamefont
  {Siepmann}(1998)}]{martin1998transferable}%
  \BibitemOpen
  \bibfield  {author} {\bibinfo {author} {\bibfnamefont {M.~G.}\ \bibnamefont
  {Martin}}\ and\ \bibinfo {author} {\bibfnamefont {J.~I.}\ \bibnamefont
  {Siepmann}},\ }\bibfield  {title} {\enquote {\bibinfo {title} {Transferable
  potentials for phase equilibria. 1. united-atom description of n-alkanes},}\
  }\href@noop {} {\bibfield  {journal} {\bibinfo  {journal} {Journal of
  Physical Chemistry B}\ }\textbf {\bibinfo {volume} {102}},\ \bibinfo {pages}
  {2569--2577} (\bibinfo {year} {1998})}\BibitemShut {NoStop}%
\bibitem [{\citenamefont {Berendsen}\ \emph {et~al.}(1981)\citenamefont
  {Berendsen}, \citenamefont {Postma}, \citenamefont {van Gunsteren},\ and\
  \citenamefont {Hermans}}]{water}%
  \BibitemOpen
  \bibfield  {author} {\bibinfo {author} {\bibfnamefont {H.}~\bibnamefont
  {Berendsen}}, \bibinfo {author} {\bibfnamefont {J.}~\bibnamefont {Postma}},
  \bibinfo {author} {\bibfnamefont {W.}~\bibnamefont {van Gunsteren}}, \ and\
  \bibinfo {author} {\bibfnamefont {J.}~\bibnamefont {Hermans}},\ }\bibfield
  {title} {\enquote {\bibinfo {title} {Interaction models for water in relation
  to protein hydration},}\ }\href@noop {} {\bibfield  {journal} {\bibinfo
  {journal} {In `Intermolecular Forces', ed. B. Pullman, Reidel, Dordrecht}\ ,\
  \bibinfo {pages} {331--342}} (\bibinfo {year} {1981})}\BibitemShut {NoStop}%
\bibitem [{\citenamefont {Allen}\ and\ \citenamefont
  {Tildesley}(1989)}]{allen1989computer}%
  \BibitemOpen
  \bibfield  {author} {\bibinfo {author} {\bibfnamefont {M.~P.}\ \bibnamefont
  {Allen}}\ and\ \bibinfo {author} {\bibfnamefont {D.~J.}\ \bibnamefont
  {Tildesley}},\ }\href@noop {} {\emph {\bibinfo {title} {Computer Simulations
  of Liquids}}}\ (\bibinfo  {publisher} {Oxford University Press},\ \bibinfo
  {year} {1989})\BibitemShut {NoStop}%
\bibitem [{\citenamefont {Essmann}\ \emph {et~al.}(1995)\citenamefont
  {Essmann}, \citenamefont {Perera}, \citenamefont {Berkowitz}, \citenamefont
  {Darden}, \citenamefont {Lee},\ and\ \citenamefont
  {Pedersen}}]{essmann1995smooth}%
  \BibitemOpen
  \bibfield  {author} {\bibinfo {author} {\bibfnamefont {U.}~\bibnamefont
  {Essmann}}, \bibinfo {author} {\bibfnamefont {L.}~\bibnamefont {Perera}},
  \bibinfo {author} {\bibfnamefont {M.~L.}\ \bibnamefont {Berkowitz}}, \bibinfo
  {author} {\bibfnamefont {T.}~\bibnamefont {Darden}}, \bibinfo {author}
  {\bibfnamefont {H.}~\bibnamefont {Lee}}, \ and\ \bibinfo {author}
  {\bibfnamefont {L.~G.}\ \bibnamefont {Pedersen}},\ }\bibfield  {title}
  {\enquote {\bibinfo {title} {A smooth particle mesh {Ewald} method},}\
  }\href@noop {} {\bibfield  {journal} {\bibinfo  {journal} {Journal of
  Chemical Physics}\ }\textbf {\bibinfo {volume} {103}},\ \bibinfo {pages}
  {8577--8593} (\bibinfo {year} {1995})}\BibitemShut {NoStop}%
\bibitem [{\citenamefont {Nos{\'e}}(1984)}]{nose1984unified}%
  \BibitemOpen
  \bibfield  {author} {\bibinfo {author} {\bibfnamefont {S.}~\bibnamefont
  {Nos{\'e}}},\ }\bibfield  {title} {\enquote {\bibinfo {title} {A unified
  formulation of the constant temperature molecular dynamics methods},}\
  }\href@noop {} {\bibfield  {journal} {\bibinfo  {journal} {Journal of
  Chemical Physics}\ }\textbf {\bibinfo {volume} {81}},\ \bibinfo {pages}
  {511--519} (\bibinfo {year} {1984})}\BibitemShut {NoStop}%
\bibitem [{\citenamefont {Parrinello}\ and\ \citenamefont
  {Rahman}(1981)}]{parrinello1981polymorphic}%
  \BibitemOpen
  \bibfield  {author} {\bibinfo {author} {\bibfnamefont {M.}~\bibnamefont
  {Parrinello}}\ and\ \bibinfo {author} {\bibfnamefont {A.}~\bibnamefont
  {Rahman}},\ }\bibfield  {title} {\enquote {\bibinfo {title} {Polymorphic
  transitions in single crystals: A new molecular dynamics method},}\
  }\href@noop {} {\bibfield  {journal} {\bibinfo  {journal} {Journal of Applied
  Physics}\ }\textbf {\bibinfo {volume} {52}},\ \bibinfo {pages} {7182--7190}
  (\bibinfo {year} {1981})}\BibitemShut {NoStop}%
\bibitem [{\citenamefont {Hockney}\ and\ \citenamefont
  {Eastwood}(2010)}]{hockney2010computer}%
  \BibitemOpen
  \bibfield  {author} {\bibinfo {author} {\bibfnamefont {R.~W.}\ \bibnamefont
  {Hockney}}\ and\ \bibinfo {author} {\bibfnamefont {J.~W.}\ \bibnamefont
  {Eastwood}},\ }\href@noop {} {\emph {\bibinfo {title} {Computer Simulation
  Using Particles}}}\ (\bibinfo  {publisher} {CRC Press},\ \bibinfo {year}
  {2010})\BibitemShut {NoStop}%
\bibitem [{\citenamefont {Honda}\ \emph {et~al.}(2004)\citenamefont {Honda},
  \citenamefont {Yamasaki}, \citenamefont {Sawada},\ and\ \citenamefont
  {Morii}}]{Honda2004}%
  \BibitemOpen
  \bibfield  {author} {\bibinfo {author} {\bibfnamefont {S.}~\bibnamefont
  {Honda}}, \bibinfo {author} {\bibfnamefont {K.}~\bibnamefont {Yamasaki}},
  \bibinfo {author} {\bibfnamefont {Y.}~\bibnamefont {Sawada}}, \ and\ \bibinfo
  {author} {\bibfnamefont {H.}~\bibnamefont {Morii}},\ }\bibfield  {title}
  {\enquote {\bibinfo {title} {10 residue folded peptide designed by segment
  statistics},}\ }\href {\doibase 10.1016/j.str.2004.05.022} {\bibfield
  {journal} {\bibinfo  {journal} {Structure (London, England : 1993)}\ }\textbf
  {\bibinfo {volume} {12}},\ \bibinfo {pages} {1507---1518} (\bibinfo {year}
  {2004})}\BibitemShut {NoStop}%
\bibitem [{\citenamefont {{Bowers}}\ \emph {et~al.}(2006)\citenamefont
  {{Bowers}}, \citenamefont {{Chow}}, \citenamefont {{Xu}}, \citenamefont
  {{Dror}}, \citenamefont {{Eastwood}}, \citenamefont {{Gregersen}},
  \citenamefont {{Klepeis}}, \citenamefont {{Kolossvary}}, \citenamefont
  {{Moraes}}, \citenamefont {{Sacerdoti}}, \citenamefont {{Salmon}},
  \citenamefont {{Shan}},\ and\ \citenamefont {{Shaw}}}]{4090217}%
  \BibitemOpen
  \bibfield  {author} {\bibinfo {author} {\bibfnamefont {K.~J.}\ \bibnamefont
  {{Bowers}}}, \bibinfo {author} {\bibfnamefont {D.~E.}\ \bibnamefont
  {{Chow}}}, \bibinfo {author} {\bibfnamefont {H.}~\bibnamefont {{Xu}}},
  \bibinfo {author} {\bibfnamefont {R.~O.}\ \bibnamefont {{Dror}}}, \bibinfo
  {author} {\bibfnamefont {M.~P.}\ \bibnamefont {{Eastwood}}}, \bibinfo
  {author} {\bibfnamefont {B.~A.}\ \bibnamefont {{Gregersen}}}, \bibinfo
  {author} {\bibfnamefont {J.~L.}\ \bibnamefont {{Klepeis}}}, \bibinfo {author}
  {\bibfnamefont {I.}~\bibnamefont {{Kolossvary}}}, \bibinfo {author}
  {\bibfnamefont {M.~A.}\ \bibnamefont {{Moraes}}}, \bibinfo {author}
  {\bibfnamefont {F.~D.}\ \bibnamefont {{Sacerdoti}}}, \bibinfo {author}
  {\bibfnamefont {J.~K.}\ \bibnamefont {{Salmon}}}, \bibinfo {author}
  {\bibfnamefont {Y.}~\bibnamefont {{Shan}}}, \ and\ \bibinfo {author}
  {\bibfnamefont {D.~E.}\ \bibnamefont {{Shaw}}},\ }\bibfield  {title}
  {\enquote {\bibinfo {title} {Scalable algorithms for molecular dynamics
  simulations on commodity clusters},}\ }in\ \href@noop {} {\emph {\bibinfo
  {booktitle} {SC '06: Proceedings of the 2006 ACM/IEEE Conference on
  Supercomputing}}}\ (\bibinfo {year} {2006})\ pp.\ \bibinfo {pages}
  {43--43}\BibitemShut {NoStop}%
\bibitem [{\citenamefont {Shaw}\ \emph {et~al.}(2010)\citenamefont {Shaw},
  \citenamefont {Maragakis}, \citenamefont {Lindorff-Larsen}, \citenamefont
  {Piana}, \citenamefont {Dror}, \citenamefont {Eastwood}, \citenamefont
  {Bank}, \citenamefont {Jumper}, \citenamefont {Salmon}, \citenamefont
  {Shan},\ and\ \citenamefont {Wriggers}}]{Shaw341}%
  \BibitemOpen
  \bibfield  {author} {\bibinfo {author} {\bibfnamefont {D.~E.}\ \bibnamefont
  {Shaw}}, \bibinfo {author} {\bibfnamefont {P.}~\bibnamefont {Maragakis}},
  \bibinfo {author} {\bibfnamefont {K.}~\bibnamefont {Lindorff-Larsen}},
  \bibinfo {author} {\bibfnamefont {S.}~\bibnamefont {Piana}}, \bibinfo
  {author} {\bibfnamefont {R.~O.}\ \bibnamefont {Dror}}, \bibinfo {author}
  {\bibfnamefont {M.~P.}\ \bibnamefont {Eastwood}}, \bibinfo {author}
  {\bibfnamefont {J.~A.}\ \bibnamefont {Bank}}, \bibinfo {author}
  {\bibfnamefont {J.~M.}\ \bibnamefont {Jumper}}, \bibinfo {author}
  {\bibfnamefont {J.~K.}\ \bibnamefont {Salmon}}, \bibinfo {author}
  {\bibfnamefont {Y.}~\bibnamefont {Shan}}, \ and\ \bibinfo {author}
  {\bibfnamefont {W.}~\bibnamefont {Wriggers}},\ }\bibfield  {title} {\enquote
  {\bibinfo {title} {Atomic-level characterization of the structural dynamics
  of proteins},}\ }\href {\doibase 10.1126/science.1187409} {\bibfield
  {journal} {\bibinfo  {journal} {Science}\ }\textbf {\bibinfo {volume}
  {330}},\ \bibinfo {pages} {341--346} (\bibinfo {year} {2010})}\BibitemShut
  {NoStop}%
\bibitem [{\citenamefont {Lindorff-Larsen}\ \emph {et~al.}(2011)\citenamefont
  {Lindorff-Larsen}, \citenamefont {Piana}, \citenamefont {Dror},\ and\
  \citenamefont {Shaw}}]{deshaw}%
  \BibitemOpen
  \bibfield  {author} {\bibinfo {author} {\bibfnamefont {K.}~\bibnamefont
  {Lindorff-Larsen}}, \bibinfo {author} {\bibfnamefont {S.}~\bibnamefont
  {Piana}}, \bibinfo {author} {\bibfnamefont {R.~O.}\ \bibnamefont {Dror}}, \
  and\ \bibinfo {author} {\bibfnamefont {D.~E.}\ \bibnamefont {Shaw}},\
  }\bibfield  {title} {\enquote {\bibinfo {title} {How fast-folding proteins
  fold},}\ }\href {\doibase 10.1126/science.1208351} {\bibfield  {journal}
  {\bibinfo  {journal} {Science}\ }\textbf {\bibinfo {volume} {334}},\ \bibinfo
  {pages} {517-- 520} (\bibinfo {year} {2011})}\BibitemShut {NoStop}%
\bibitem [{\citenamefont {Piana}, \citenamefont {Lindorff-Larsen},\ and\
  \citenamefont {Shaw}(2011)}]{PMID:21539772}%
  \BibitemOpen
  \bibfield  {author} {\bibinfo {author} {\bibfnamefont {S.}~\bibnamefont
  {Piana}}, \bibinfo {author} {\bibfnamefont {K.}~\bibnamefont
  {Lindorff-Larsen}}, \ and\ \bibinfo {author} {\bibfnamefont {D.~E.}\
  \bibnamefont {Shaw}},\ }\bibfield  {title} {\enquote {\bibinfo {title} {How
  robust are protein folding simulations with respect to force field
  parameterization?}}\ }\href {\doibase 10.1016/j.bpj.2011.03.051} {\bibfield
  {journal} {\bibinfo  {journal} {Biophysical Journal}\ }\textbf {\bibinfo
  {volume} {100}},\ \bibinfo {pages} {L47---9} (\bibinfo {year}
  {2011})}\BibitemShut {NoStop}%
\bibitem [{\citenamefont {Jorgensen}\ \emph {et~al.}(1983)\citenamefont
  {Jorgensen}, \citenamefont {Chandrasekhar}, \citenamefont {Madura},
  \citenamefont {Impey},\ and\ \citenamefont
  {Klein}}]{jorgensen1983comparison}%
  \BibitemOpen
  \bibfield  {author} {\bibinfo {author} {\bibfnamefont {W.~L.}\ \bibnamefont
  {Jorgensen}}, \bibinfo {author} {\bibfnamefont {J.}~\bibnamefont
  {Chandrasekhar}}, \bibinfo {author} {\bibfnamefont {J.~D.}\ \bibnamefont
  {Madura}}, \bibinfo {author} {\bibfnamefont {R.~W.}\ \bibnamefont {Impey}}, \
  and\ \bibinfo {author} {\bibfnamefont {M.~L.}\ \bibnamefont {Klein}},\
  }\bibfield  {title} {\enquote {\bibinfo {title} {Comparison of simple
  potential functions for simulating liquid water},}\ }\href@noop {} {\bibfield
   {journal} {\bibinfo  {journal} {Journal of Chemical Physics}\ }\textbf
  {\bibinfo {volume} {79}},\ \bibinfo {pages} {926--935} (\bibinfo {year}
  {1983})}\BibitemShut {NoStop}%
\bibitem [{\citenamefont {Shan}\ \emph {et~al.}(2005)\citenamefont {Shan},
  \citenamefont {Klepeis}, \citenamefont {Eastwood}, \citenamefont {Dror},\
  and\ \citenamefont {Shaw}}]{doi:10.1063/1.1839571}%
  \BibitemOpen
  \bibfield  {author} {\bibinfo {author} {\bibfnamefont {Y.}~\bibnamefont
  {Shan}}, \bibinfo {author} {\bibfnamefont {J.~L.}\ \bibnamefont {Klepeis}},
  \bibinfo {author} {\bibfnamefont {M.~P.}\ \bibnamefont {Eastwood}}, \bibinfo
  {author} {\bibfnamefont {R.~O.}\ \bibnamefont {Dror}}, \ and\ \bibinfo
  {author} {\bibfnamefont {D.~E.}\ \bibnamefont {Shaw}},\ }\bibfield  {title}
  {\enquote {\bibinfo {title} {Gaussian split ewald: A fast ewald mesh method
  for molecular simulation},}\ }\href {\doibase 10.1063/1.1839571} {\bibfield
  {journal} {\bibinfo  {journal} {The Journal of Chemical Physics}\ }\textbf
  {\bibinfo {volume} {122}},\ \bibinfo {pages} {054101} (\bibinfo {year}
  {2005})}\BibitemShut {NoStop}%
\bibitem [{\citenamefont {Hoover}(1985)}]{Hoover}%
  \BibitemOpen
  \bibfield  {author} {\bibinfo {author} {\bibfnamefont {W.~G.}\ \bibnamefont
  {Hoover}},\ }\bibfield  {title} {\enquote {\bibinfo {title} {Canonical
  dynamics: Equilibrium phase-space distributions},}\ }\href {\doibase
  10.1103/PhysRevA.31.1695} {\bibfield  {journal} {\bibinfo  {journal}
  {Physical Review A}\ }\textbf {\bibinfo {volume} {31}},\ \bibinfo {pages}
  {1695--1697} (\bibinfo {year} {1985})}\BibitemShut {NoStop}%
\bibitem [{\citenamefont {Kingma}\ and\ \citenamefont
  {Ba}(2014)}]{kingma2014adam}%
  \BibitemOpen
  \bibfield  {author} {\bibinfo {author} {\bibfnamefont {D.~P.}\ \bibnamefont
  {Kingma}}\ and\ \bibinfo {author} {\bibfnamefont {J.}~\bibnamefont {Ba}},\
  }\href@noop {} {\enquote {\bibinfo {title} {Adam: A method for stochastic
  optimization},}\ } (\bibinfo {year} {2014}),\ \Eprint
  {http://arxiv.org/abs/1412.6980} {arXiv:1412.6980} \BibitemShut {NoStop}%
\bibitem [{\citenamefont {Mittal}\ and\ \citenamefont
  {Shukla}(2018)}]{Mittal2018}%
  \BibitemOpen
  \bibfield  {author} {\bibinfo {author} {\bibfnamefont {S.}~\bibnamefont
  {Mittal}}\ and\ \bibinfo {author} {\bibfnamefont {D.}~\bibnamefont
  {Shukla}},\ }\bibfield  {title} {\enquote {\bibinfo {title} {Maximizing
  kinetic information gain of markov state models for optimal design of
  spectroscopy experiments},}\ }\href {\doibase 10.1021/acs.jpcb.8b07076}
  {\bibfield  {journal} {\bibinfo  {journal} {The Journal of Physical Chemistry
  B}\ }\textbf {\bibinfo {volume} {122}},\ \bibinfo {pages} {10793--10805}
  (\bibinfo {year} {2018})}\BibitemShut {NoStop}%
\bibitem [{\citenamefont {Pietrucci}(2020)}]{pietrucci2020novel}%
  \BibitemOpen
  \bibfield  {author} {\bibinfo {author} {\bibfnamefont {F.}~\bibnamefont
  {Pietrucci}},\ }\bibfield  {title} {\enquote {\bibinfo {title} {Novel
  enhanced sampling strategies for transitions between ordered and disordered
  structures},}\ }\href@noop {} {\bibfield  {journal} {\bibinfo  {journal}
  {Handbook of Materials Modeling: Methods: Theory and Modeling}\ ,\ \bibinfo
  {pages} {597--619}} (\bibinfo {year} {2020})}\BibitemShut {NoStop}%
\bibitem [{\citenamefont {Han}\ \emph {et~al.}(2018)\citenamefont {Han},
  \citenamefont {Zhang}, \citenamefont {Car},\ and\ \citenamefont
  {E}}]{Han_2018}%
  \BibitemOpen
  \bibfield  {author} {\bibinfo {author} {\bibfnamefont {J.}~\bibnamefont
  {Han}}, \bibinfo {author} {\bibfnamefont {L.}~\bibnamefont {Zhang}}, \bibinfo
  {author} {\bibfnamefont {R.}~\bibnamefont {Car}}, \ and\ \bibinfo {author}
  {\bibfnamefont {W.}~\bibnamefont {E}},\ }\bibfield  {title} {\enquote
  {\bibinfo {title} {Deep potential: A general representation of a many-body
  potential energy surface},}\ }\href@noop {} {\bibfield  {journal} {\bibinfo
  {journal} {Communications in Computational Physics}\ }\textbf {\bibinfo
  {volume} {23}} (\bibinfo {year} {2018})}\BibitemShut {NoStop}%
\end{thebibliography}%

\end{document}